\documentclass{aa}  
\usepackage{graphicx}
\usepackage{subfigure}
\usepackage{hyperref}
\hypersetup{
    colorlinks = true,
    citecolor ={blue}
}
\usepackage{txfonts}
\usepackage{verbatim}
\usepackage{soul}
\usepackage{float}
\usepackage{pifont}
\usepackage{tabularx,adjustbox,booktabs}
\begin{document}

\title{Pulsating hydrogen-deficient white dwarfs and pre-white dwarfs 
observed with {\it TESS}} 
\subtitle{III. Asteroseismology of the DBV star GD~358}

\author{Alejandro H. C\'orsico\inst{1,2}, 
        Murat Uzundag\inst{3,4},   
        S. O. Kepler\inst{5}, 
        Roberto Silvotti\inst{6},
        Leandro G. Althaus\inst{1,2},
        Detlev Koester\inst{7},
        Andrzej S. Baran\inst{8,9,10}, 
        Keaton J. Bell\inst{11,12},
        Agn\`es Bischoff-Kim\inst{13},
        J. J. Hermes\inst{14},
        Steve D. Kawaler\inst{15}, 
        Judith L. Provencal\inst{16,17},
        Don E. Winget\inst{18,19}, 
        Michael H. Montgomery\inst{18,19}, 
        Paul A. Bradley\inst{20},
        S. J. Kleinman\inst{21}, and
        Atsuko Nitta\inst{21}
        }
\institute{Grupo de Evoluci\'on Estelar y Pulsaciones. 
           Facultad de Ciencias Astron\'omicas y Geof\'{\i}sicas, 
           Universidad Nacional de La Plata, 
           Paseo del Bosque s/n, 1900 
           La Plata, 
           Argentina
           \and
           IALP - CONICET
           \and
           Instituto de F\'isica y Astronom\'ia, Universidad de Valparaiso, Gran Breta\~na 1111, Playa Ancha, Valpara\'iso 2360102, Chile
           \and
           European Southern Observatory, Alonso de Cordova 3107, Santiago, Chile
           \and
           Instituto de F\'{i}sica, Universidade Federal do Rio Grande do Sul, 91501-970, Porto-Alegre, RS, Brazil
           \and 
           INAF-Osservatorio Astrofisico di Torino, strada dell'Osservatorio 20, 10025 Pino Torinese, Italy
           \and 
           Institut f\"ur Theoretische Physik und Astrophysik, Universit\"at Kiel, 24098 Kiel, Germany
           \and
           ARDASTELLA Research Group, Institute of Physics, Pedagogical University of Krakow, ul. Podchor\c{a}\.zych 2, 30-084 Krak\'ow, Poland
           \and
           Embry-Riddle Aeronautical University, Department of Physical Science, Daytona Beach, FL\,32114, USA
           \and
           Department of Physics, Astronomy, and Materials Science, Missouri State University, Springfield, MO\,65897, USA
           \and
           DIRAC Institute, Department of Astronomy, University of Washington, Seattle, WA-98195, USA 
           \and
           NSF Astronomy and Astrophysics Postdoctoral Fellow
           \and
           Penn State Worthington Scranton, Dunmore, PA 18512, USA
           \and
           Department of Astronomy, Boston University, Boston, MA-02215, USA
           \and
           Department of Physics and Astronomy, Iowa State University, Ames, IA 50011, USA
           \and
           University  of  Delaware,  Department  of  Physics  and  Astronomy  Newark,  DE  19716, USA
           \and
           Delaware Asteroseismic Research Center, Mt.  Cuba Observatory, Greenville, 
           DE 19807, USA
           \and
           Department of Astronomy, University of Texas at Austin, Austin, TX-78712, USA 
           \and
           McDonald Observatory, Fort Davis, TX-79734, USA
           \and
           XCP-6, MS F-699 Los Alamos National Laboratory, Los Alamos, NM 87545, USA
           \and
           Gemini Observatory/NSF’s NOIRLab, 670 N. A’ohoku Place, Hilo, Hawai’i, 96720, USA
           }
           
           
\date{Received ; accepted }
\abstract{
The collection of high-quality photometric data by space telescopes, such as the 
already completed {\it Kepler} mission and the ongoing {\it TESS}  program, 
is revolutionizing the area of white-dwarf asteroseismology. Among the different kinds of 
pulsating white dwarfs, there are those that have He-rich atmospheres, and they are called 
DBVs or V777 Her variable stars. The archetype of these pulsating white dwarfs, GD~358, is 
the focus of the present paper.}
{We report a thorough asteroseismological analysis of the DBV star GD~358 (TIC~219074038) 
based on new high-precision photometric data gathered by the {\it TESS} space mission 
combined with data taken from the Earth.}
{We reduced {\it TESS} observations of the DBV star GD~358 and performed a 
detailed asteroseismological analysis using fully evolutionary DB white-dwarf models 
computed accounting the complete 
prior evolution of their progenitors. We assessed  the  mass  of this  star  
by comparing  the  measured  mean period separation with the theoretical  averaged 
period spacings of the models, and we used the observed individual periods to look for 
a seismological stellar model. We detected potential frequency multiplets 
for GD~358, which we use to identify the harmonic degree ($\ell$) of the pulsation 
modes and rotation period.}
{In total, we detected 26 periodicities from the {\it TESS} light
  curve of this DBV star using a standard  pre-whitening. The
  oscillation frequencies are associated with nonradial $g$(gravity)-mode
  pulsations with periods from  $\sim 422$ s to $\sim 1087$ s. Moreover, 
  we detected 8 combination frequencies between $\sim 543$ s and $\sim 295$ s.  
  We combined these  data with a huge amount of observations from the ground. 
  We found a constant period spacing of $39.25\pm0.17$ s, which helped us to 
  infer its mass ($M_{\star}= 0.588\pm0.024 M_{\sun}$) and constrain 
  the  harmonic degree $\ell$ of the
  modes. We carried out a period-fit analysis on  GD~358, and we were successful 
  in finding an asteroseismological model with a stellar mass ($M_{\star}= 
  0.584^{+0.025}_{-0.019} M_{\sun}$), compatible with the stellar mass derived 
  from the period spacing,  and in line with the spectroscopic mass ($M_{\star}= 
  0.560\pm0.028 M_{\sun}$). In agreement with previous works, 
  we found that the frequency splittings vary according to the radial order of 
  the modes, suggesting differential rotation.
  Obtaining a seismological made it possible to estimate the seismological 
  distance ($d_{\rm seis}= 42.85\pm 0.73$ pc) of 
  GD~358, which is in  very good  accordance with the precise astrometric  
  distance measured by 
  {\it GAIA} EDR3 ($\pi= 23.244\pm 0.024, d_{\rm GAIA}= 43.02\pm 0.04$~pc).}   
{The high-quality of data measured with the {\it TESS} space telescope, used in 
combination with data taken from ground-based observatories, provides invaluable 
information for conducting asteroseismological studies of DBV  stars, analogously 
to what happens with other types  of pulsating white-dwarf stars. The currently 
operating {\it TESS} mission, together with the advent of other similar space 
missions and new stellar surveys, will give an unprecedented boost to white 
dwarf asteroseismology.}

\authorrunning{C\'orsico et al.}
\titlerunning{Asteroseismology of GD~358 with {\it TESS}}
\maketitle

\section{Introduction}

Pulsating white dwarfs (WD) and pre-WDs constitute a long-studied and
reliably established class of compact variable stars, both from observational
and theoretical grounds. Their brightness variations are
multiperiodic, with periods between $100$ s and $7\,000$ s,
and amplitudes up to $0.4$ mag in typical optical light curves  
\citep[][]{2008ARA&A..46..157W,2008PASP..120.1043F, 
2010A&ARv..18..471A, 2019A&ARv..27....7C}. The variability is 
associated to low-degree
($\ell \leq 2$) nonradial $g$(gravity)-mode pulsations 
excited by a physical mechanism related to the partial ionization of the
dominant chemical species in the zone of driving,
located in the external layers. In warm and cool pulsating WDs, the opacity bumps associated to these 
partial-ionization zones are responsible
for the appearance of an outer convection zone, which also strongly
contributes to  $g$-mode pulsation instabilities
\citep{1991MNRAS.251..673B,1999ApJ...511..904G,1999ApJ...519..783W}.
The first pulsating WD was discovered in 1968
\citep{1968ApJ...153..151L}, and currently more than 300 objects
are known. They are classified in various categories, including 
DAVs or ZZ Ceti stars (pulsating WDs with H atmospheres), DBV or V777 Her stars 
(pulsating WDs with helium atmospheres), and pulsating PG~1159 or GW Vir
stars, among others \citep[][]{
2008ARA&A..46..157W,2008PASP..120.1043F,2010A&ARv..18..471A,
2019A&ARv..27....7C}. 

Since the discovery of the first pulsating WDs, observations of these
variable stars have been steadily increasing, thanks to
single-site observations and also multi-site campaigns like those of
the Whole Earth Telescope \citep[{\it WET};][]{1990ApJ...361..309N}.
Subsequently, a dramatic growth in the number of known pulsators 
was made possible thanks to the 
identification of candidates from the spectral observations of the
Sloan Digital Sky Survey \citep[SDSS,][]{2000AJ....120.1579Y,Kleinman13,Kepler15,Kepler16,Kepler19}, 
examples of which are the works of \cite{2004ApJ...607..982M,2004ApJ...612.1052M,2006ApJ...640..956M}.
Finally, in recent years, the area has received a strong boost driven by
the uninterrupted observations from space made by the {\it Kepler} telescope,
both main mission \citep{2010Sci...327..977B} and {\it K2} mode
\citep{2014PASP..126..398H}. Indeed, these efforts paved the 
way to the analysis of 32 ZZ Ceti stars and three DBV stars \citep[][]{2011ApJ...736L..39O,2014MNRAS.438.3086G, 2015ApJ...809...14B,2017ApJS..232...23H,2017ApJ...835..277H,
  2017ApJ...851...24B,2017PhDT........14C,2021arXiv210813988D}, 
  until the mission was terminated due to lack of fuel in 2018. The successor to {\it Kepler} is the Transiting Exoplanet Survey
Satellite \citep[{\it TESS},][]{2015JATIS...1a4003R}. This space mission 
has observed 200\,000 brightest stars in 85\% of the whole sky in
2019 and 2020 in the first part of the mission. {\it TESS}
performs extensive time-series photometry that allows to
discover pulsating stars, and, in particular, variable hot subdwarfs, WDs,
and pre-WDs  with mag < 16, with short (120\,s) cadence. Starting in July 2020,
it is now also observing in 20\,s cadence.

Relevant to this work are the DBV  stars, which are  
pulsating He atmosphere WDs with effective  temperatures  in
the range $22\,400 \lesssim T_{\rm eff} \lesssim 32\,000$~K and pulsate with $g$-mode 
periods between 120 and 1080~s
\citep[][]{2008ARA&A..46..157W,2019A&ARv..27....7C}. The existence of the
DBV class of pulsating WDs was anticipated through theoretical
arguments \citep{1982ApJ...252L..65W} before it was confirmed observationally
shortly after \citep{1982ApJ...262L..11W}. Pulsations in DBVs
are thought to be excited by a combination of the $\kappa$ mechanism acting in
the He partial ionization zone --- and thus setting the blue edge of the DBV
instability strip \citep{1983ApJ...268L..33W},  and  the ``convective driving''
mechanism \citep{1991MNRAS.251..673B,2008JPhCS.118a2051D,2008ASPC..391..183Q,2017ASPC..509..321V}.

Due to the high-quality {\it Kepler} and {\it K2}
observations, three DBV stars, KIC~8626021 \citep{2011ApJ...736L..39O},
PG~0112+104 \citep{2017ApJ...835..277H}, and EPIC 228782059 \citep{2021arXiv210813988D}, were
intensively studied with space data. In particular, KIC~8626021 has 
been repeatedly modelled by several independent research groups
\citep{2011ApJ...742L..16B,2012A&A...541A..42C,2014ApJ...794...39B,
2018Natur.554...73G,2019A&A...628L...2C}, who have explored its
internal structure with unprecedented precision.  The first DBV pulsator 
observed extensively with {\it TESS}, the star EC~0158$-$160 or WD~0158$-$160 (TIC~257459955), was  analyzed and modelled by \cite{2019A&A...632A..42B}, who 
found nine independent frequencies appropriate for asteroseismology. 

In this work, we present new {\it TESS} observations of the already known DBV star 
GD~358.  It has been scrutinized extensively from the ground for 3 
decades. While some time-series photometry of GD~358 has been obtained from
space \citep[e.g.,][]{2005A&A...432..175C}, in this work we examine this
archetypal DBV star using intensive high-precision photometry from space for the 
first time. We also perform a detailed asteroseismological analysis of 
this star on the basis of the fully evolutionary models of DB WDs computed 
by \cite{2009ApJ...704.1605A}. The present study is the third part of our series of 
papers devoted to the study of pulsating H-deficient
WDs observed with {\it TESS}. The first article is focused
on six already known GW Vir stars \citep{2021A&A...645A.117C}, and the second  
one is devoted to the discovery of two new GW Vir stars, specifically DOVs \citep{2021A&A...655A..27U}.

The paper is  organized as  follows.  In Section \ref{targets} 
we provide a brief  account of the main characteristics of
GD~358. In  Sect. \ref{observations}, we
describe the methods we apply to obtain the pulsation periods of the
target star.
A brief summary of the stellar
models of DB WD stars employed for the asteroseismological analysis
of GD~358 is  provided  in Sect.  \ref{models}.
Sect. \ref{asteroseismology} is devoted to the asteroseismological modelling 
of the target star, including the look for a possible uniform 
period spacing in the period spectrum by
applying significance tests, the derivation of the stellar mass 
using the period separation, and by carrying out a
period-to-period fit with the goal of finding an asteroseismological
model.  Finally, in Sect. \ref{conclusions}, we summarize our results.

\begin{figure} 
\includegraphics[clip,width=1.0\columnwidth]{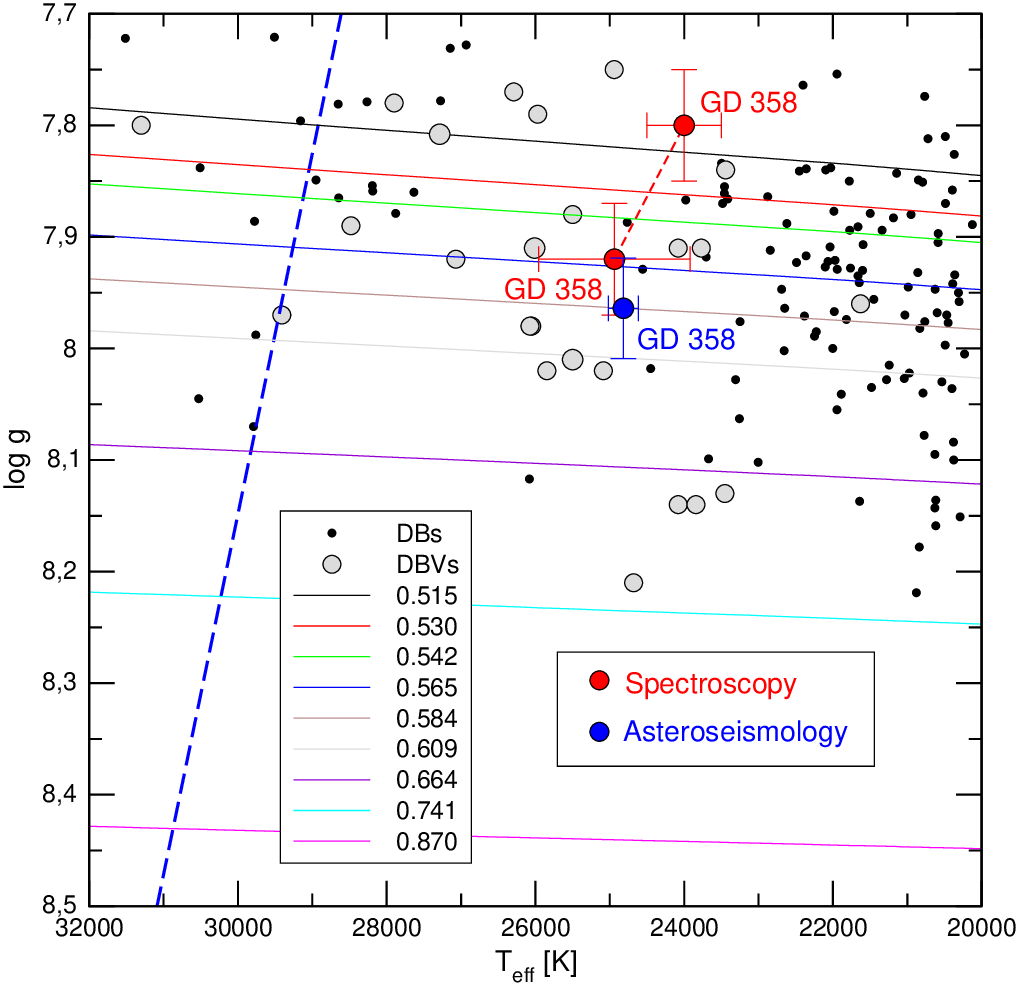}
\caption{Location  of DB WDs on the $T_{\rm eff}-\log g$ diagram \citep{2019MNRAS.486.2169K}, marked with small black circles. Also depicted is the location of the published  DBV  stars
(gray circles), according  to  the  compilation  by  \cite{2019A&ARv..27....7C}. 
The target star of the present paper, GD~358, is highlighted with 
large red circles according to spectroscopy, being the two locations of 
this star corresponding to two spectroscopic determinations of $T_{\rm eff}$ and $\log g$ 
according  to \cite{2014A&A...568A.118K} ($T_{\rm eff}= 24\,000\pm500$ K, $\log g= 7.80\pm0.05$), which are very close to the 
recent derivations by \cite{Kong2021}, and according to 
\cite{2017ApJ...848...11B} ($T_{\rm eff}= 24\,937\pm1018$ K, $\log g= 7.92\pm0.05$). 
The location
of the asteroseismological model (see Sect. \ref{p-to-p-fits-GD358}) 
is emphasized with a blue circle. The DB WD evolutionary tracks of 
\cite{2009ApJ...704.1605A} are displayed with different colors according 
to the stellar-mass values (in solar units). The blue-dashed line 
represents the theoretical dipole ($\ell= 1$) blue edge of the DBV instability 
strip, according to \cite{2009JPhCS.172a2075C}.}
\label{fig:1} 
\end{figure}

\section{The target star: GD~358}
\label{targets}

\begin{table*}
\centering
  \caption{Characteristics of GD~358. Columns 1, 2, 3, 4, 5, and 6 correspond to
  the {\it TESS} input catalog number, name of the object, effective temperature, surface gravity,  
  Gaia EDR3 parallax, and distance, respectively.  There are three 
  spectroscopic determinations of the atmospheric parameters, the first row
  corresponding to the values from \cite{2017ApJ...848...11B}, the second one from \cite{2012ASPC..462..171N,2014A&A...568A.118K}, and the third one from \cite{Kong2021}. For details, see the text.}
  \begin{tabular}{cccccc}
\hline
TIC        & Name             &  $T_{\rm eff}$ & $\log g$  & $\pi$   &  $d$  \\
           &                  &    [K]         & [cgs]     & [mas]   &   [pc]\\
\hline
219074038 & GD~358 (V777 Her) & $24\,937\pm1018$ & $7.92\pm0.05$   & $23.244\pm 0.024$ & $43.02\pm0.04$\\ 
          &                   & $24\,000\pm500$ & $7.80\pm0.05$    &                  &                 \\ 
          & LAMOST~J164718.35+322832.9  & $24\,075\pm124$ & $7.827\pm0.01$  &                  &                 \\ 
\hline
\label{basic-parameters-targets}
\end{tabular}
\end{table*}

The location of GD~358 in the $\log T_{\rm eff}$-$\log g$ diagram is depicted in 
Fig. \ref{fig:1}.  We have included the evolutionary tracks of DB WDs computed by  \cite{2009ApJ...704.1605A}.  Independently of the precise location of GD~358 
dictated by the spectroscopic parameters (see below), the star appears to be in 
the middle of the DBV instability strip, with a stellar mass somewhat lower than the average mass of the 
C/O-core WDs ($\sim 0.6 M_{\odot}$). We describe the basic characteristics 
of GD~358 below and summarize its stellar properties 
in Table \ref{basic-parameters-targets}. GD~358 (or V777 Her) has a 
{\it TESS} Input Catalog (TIC) number  TIC~219074038. It is the brightest ($m_{\rm V}=13.7$) 
and most extensively studied DBV star. This prototypical object provides the 
designation for the class of V777 Her (or DBV) variable stars. 
The pulsations of GD~358 were discovered in 1982 \citep{1982ApJ...252L..65W}. Its spectroscopic surface 
parameters are $T_{\rm eff}= 24\,937\pm 1018$ K and $\log g= 7.92\pm0.05$ according to \citet{2017ApJ...848...11B} from optical data (see Fig.\ref{fig:1}), although the previous analysis by \cite{2012ASPC..462..171N} and \cite{2014A&A...568A.118K} using optical and UV data 
give $T_{\rm eff}= 24\,000\pm500$ K and 
$\log g= 7.8\pm 0.05$ (Fig.\ref{fig:1}). 
Recently, \citet{Kong2021} derived the atmospheric parameters of 
GD~358 with LAMOST data and found $T_{\rm eff}$ = $24\,075\pm124$\,K and
$\log g = 7.827\pm0.01$\,dex. These values are in perfect agreement with the 
parameters derived by  \cite{2012ASPC..462..171N} and 
\cite{2014A&A...568A.118K}. The spectroscopic parameters 
are summarized in Table \ref{basic-parameters-targets}. 
GD~358 has been extensively observed by the {\it WET} collaboration \citep{1994ApJ...430..839W, 2000MNRAS.314..689V, 2003A&A...401..639K,2009ApJ...693..564P}. The most recent 
and complete analysis of this star has been carried out by \cite{2019ApJ...871...13B}, 
who collected and reduced data from 34 years of photometric observations, including 
archival data from 1982 to 2006, and 1195.2 hr of observations from 2007 to 2016. \cite{2019ApJ...871...13B} detected a total of 15 independent periods, of which 
thirteen belong to a series of $\ell= 1$ pulsation periods with consecutive radial 
order, the longest continuous sequence of periods observed in a DBV star until then. 
The star has repeatedly been the focus of asteroseismological analyses 
\citep{1994ApJ...430..850B,2000ApJ...545..974M, 2001ApJ...557.1021M, 2002ApJ...581L..33F, 2003ApJ...587L..43M}  
using evolutionary models of DB WDs with simplified 
chemical profiles. \cite{2019ApJ...871...13B} analyzed this star using models 
that include parameterized, complex, core-composition profiles to fit the 15 
observed periods. They obtain a seismological model with a thickness of the He layer that 
is qualitatively consistent with the diffusion-calculation picture that predict that 
the pure-He envelope will steadily grow thicker as the DB star cools \citep{1995ApJ...445L.141D, 2002ApJ...581L..33F, 2004A&A...417.1115A}.
The {\it Gaia} EDR3 parallax and distance for GD~358 are $\pi= 23.244\pm0.024$~mas and $d= 43.02 \pm 0.04$ pc \citep{Bailer-Jones2021}, respectively. 

\begin{figure*} 
\includegraphics[clip,width=1.\linewidth]{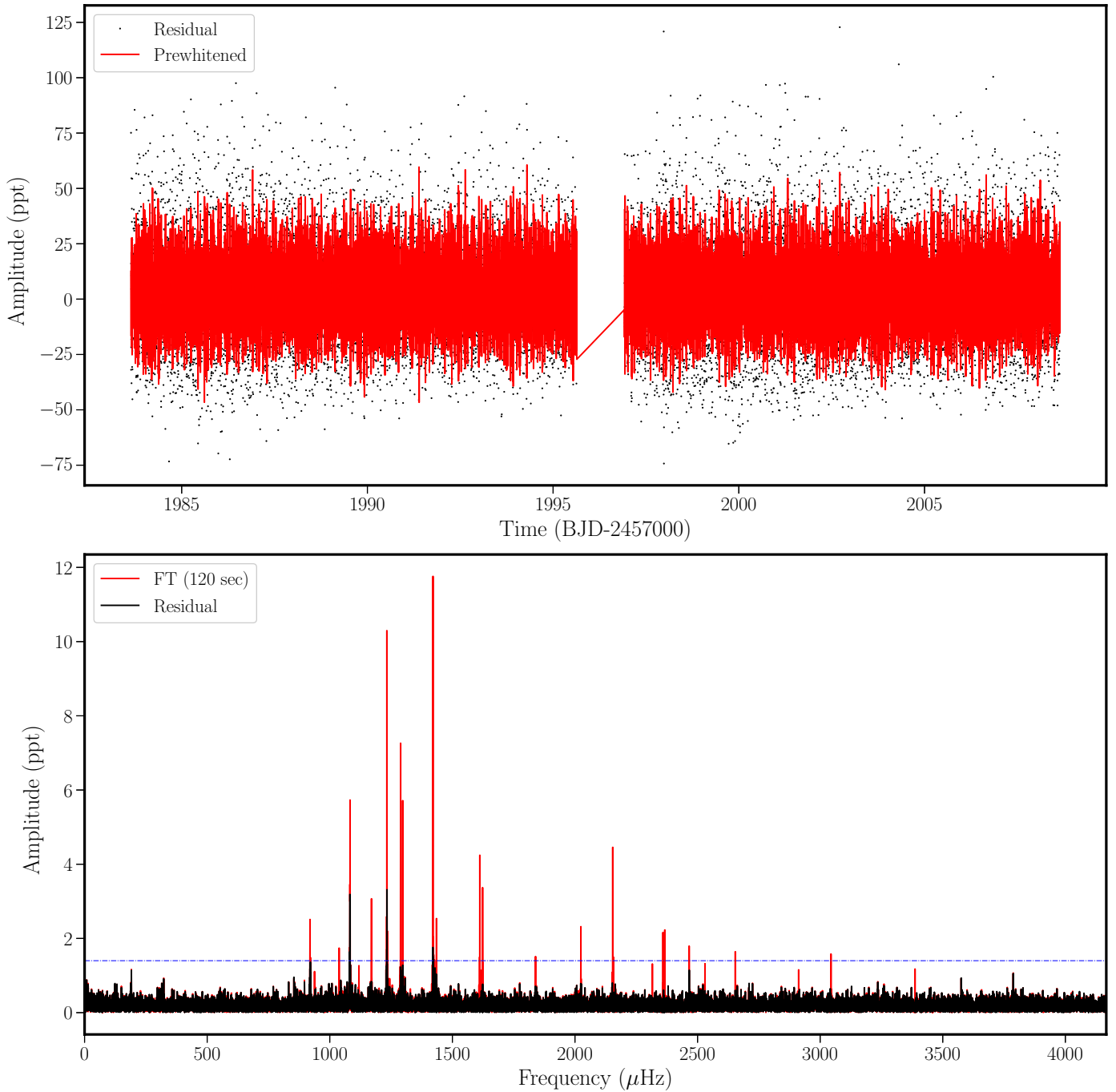}
\caption{{\sc TOP:} The light curve of the pulsating DBV star GD~358 observed in sector 25. 
The black dots show the residual
flux while the red lines show all prewhitened variations from the light curve.
{\sc BOTTOM:} Fourier transform of the pulsating DBV star GD~358 observed with 120-sec cadence. 
The dotted horizontal blue line indicates the 0.1\% false-alarm-probability (FAP) significance threshold. 
The black line is the FT of the prewhitened light curve.}
\label{fig:lightcurves_FT} 
\end{figure*}

\section{Observations and data reduction}  
\label{observations}  

GD~358 was observed by {\it TESS} at 2-min cadence on Sector 25 between 2020-May-13 and 2020-June-08, targeted as TIC\,219074038 ($T=13.9$\,mag). The temporal resolution is $1/T = 0.451\ \mu$Hz ($T$ being the data span of
25.67\,days). The light curve for GD~358 released by the Science Processing Operations Center (SPOC) pipeline had large gaps, which we believe were caused by unnecessarily harsh clipping based on quality flags that resulted in a low (67.4\%) duty cycle. We re-reduced the Sector 25 data with the same 5-pixel aperture from the SPOC pipeline but with a looser quality-flag cuts, yielding a significantly higher (92.6\%) duty cycle. Our final light curve was flattened of long-term trends by the division of second-order polynomial every 2 days.
The final light curve is shown in Fig. \ref{fig:lightcurves_FT} including 16781 data points spanning 25 days. The average noise level of the amplitude spectra is 0.24 ppt. 
We calculated a detection threshold of 0.1\%\ false alarm probability (FAP) following the method described in \citet{kepler1993}  \citep[see also][]{2021AcA....71..113B}.
If the amplitude of a given peak is above this value, there is a 0.1\% chance of resulting from random noise fluctuations.
All frequencies above the threshold level of $0.1\%$\,FAP of 1.45 ppt have been prewhitened a few exceptions. 
The frequencies at 1657 and 2157 $\mu$Hz are below the significance threshold with S/N of 5.6 and 4.6 respectively. These frequencies were reported in the previous work by \citet{2019ApJ...871...13B}. Moreover, we prewhitened 8 combination frequencies, which are located at beyond 2300 $\mu$Hz. 
A few of them at 2315, 2530, 2912 and 3386 $\mu$Hz are slightly below  $0.1\%$\,FAP level with S/N of 5.3, 5.5, 4.7 and 5 respectively. All combination frequencies beyond 2400 $\mu$Hz are detected for the first time and also extracted from the light curve and reported in Table \ref{table:GD358-murat}. 
Overall, we have detected 26 frequencies that are concentrated between 900 $\mu$Hz and 3400 $\mu$Hz, out of which we identified 8 combination frequencies.
Fig. \ref{fig:lightcurves_FT} displays the Fourier transform of  GD~358. In Table \ref{table:GD358-murat} we
show the list of periods of GD~358 detected with {\it TESS}.

\begin{figure} 
\includegraphics[clip,width=1.0\columnwidth]{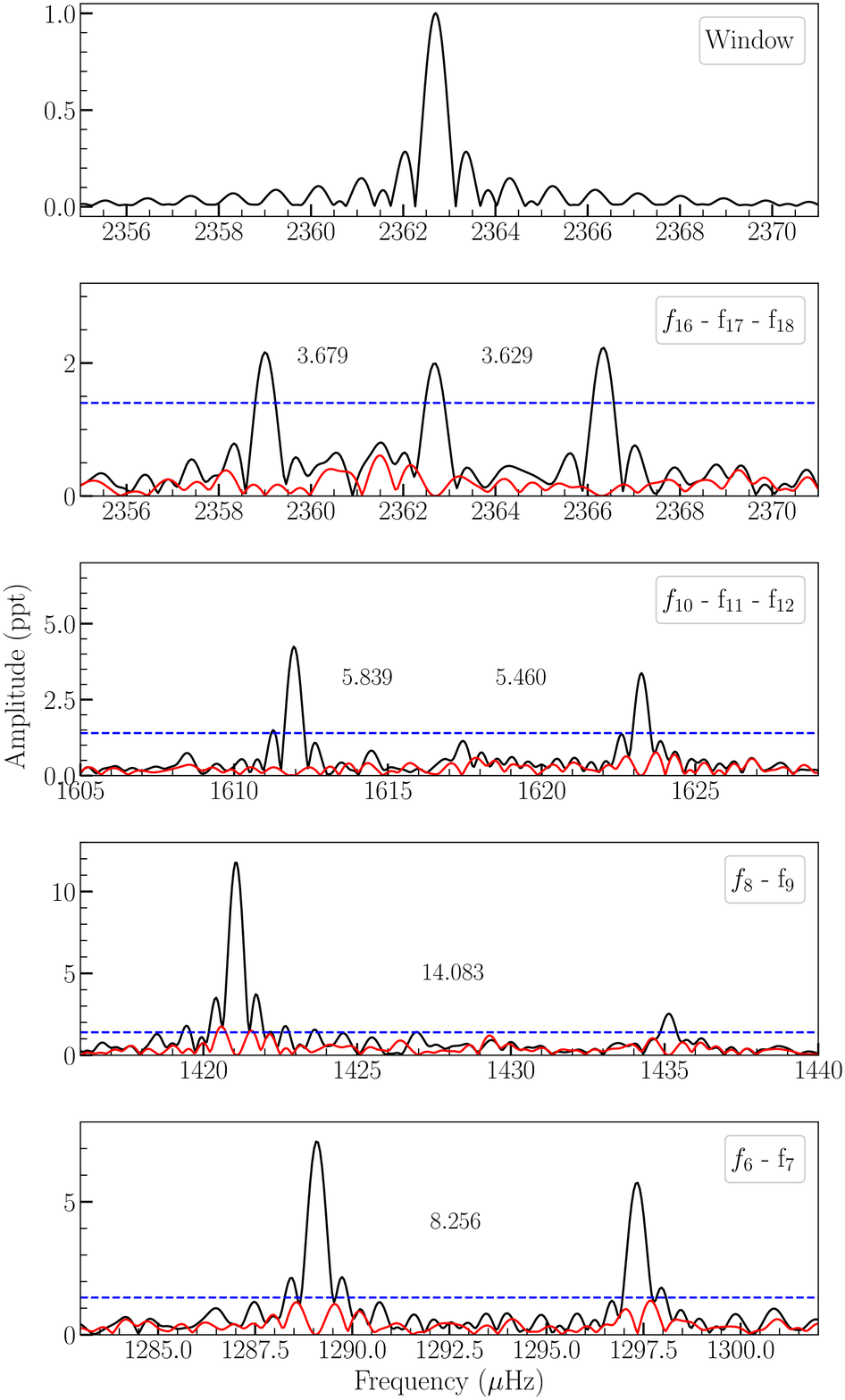}
\caption{Rotational triplets in the frequency spectrum of GD~358. The FT are shown with black lines, and residuals with red lines. The blue-dashed horizontal lines correspond to $0.1\%$ FAP confidence levels. The upper panel is the window function.}
\label{fig:RM-GD358} 
\end{figure}

The frequencies emphasized with boldface in  Table \ref{table:GD358-murat} are components of rotational triplets ($\ell= 1$). 
Rotational multiplets are depicted in Fig. \ref{fig:RM-GD358}. 
A well-known property of nonradial stellar pulsations is that the eigenfrequencies
of degree $\ell$ split into $2{\ell}+1$ components differing in azimuthal ($m$) 
number due to stellar rotation. When the rotation is slow and rigid, 
the frequency splitting can be obtained as  
$\delta \nu_{\ell, k, m}= m\ (1-C_{\ell,k})\ \Omega_{\rm
  R}$, $\Omega_{\rm R}$ being the rotational angular frequency  of the
pulsating star, and $m= 0, \pm 1, \pm 2, \cdots, \pm \ell$ \citep{1989nos..book.....U}. 
The condition of slow rotation implies that $\Omega_{\rm R} \ll
\nu_{\ell, k}$. The $C_{\ell, k}$ constants are the Ledoux coefficients \citep{1958HDP....51..353L}, that can be assessed as $C_{\ell,k} \sim [\ell(\ell+1)]^{-1}$
in the asymptotic limit of high radial-order $g$ modes ($k \gg \ell$).  
In the specific case of dipole ($\ell= 1$) and
quadrupole ($\ell= 2$) modes, we have  $C_{1,k}\sim  0.5$ and $C_{2,k}
\sim 0.17$, respectively. Besides allowing an estimate of the rotation speed of 
the star, multiplets in the frequency spectrum of a pulsating WD 
are very useful to identify the harmonic degree of the pulsation modes. 
This approach to derive the rotation period has
been successfully applied to several pulsating WD stars
\citep[see, for instance,][for the case of ZZ Ceti 
stars observed during the {\it Kepler} and 
{\it K2} missions]{2017ApJS..232...23H}. \cite{1994ApJ...430..839W} detected 10 
complete (that is, with all the three components) rotational triplets of frequencies in GD~358 as a result of an intensive 
scrutiny of this star with the {\it WET} collaboration. These authors found that 
the frequency splittings ($\delta \nu$) are not constant, 
but vary with the radial order, which led them to conclude that the star 
could be experiencing differential rotation, with the outer envelope 
rotating twice as fast as the core \citep[but see also][]{1999ApJ...516..349K}. \cite{2003A&A...401..639K} 
reported the  absence of triplets in the 2000 data of GD~358, except for one clear triplet centered at $\sim 2154\ \mu$Hz. Later, \cite{2009ApJ...693..564P}
found only two clear rotational triplets, centered at $\sim 2154\ \mu$Hz and $\sim 2363\ \mu$Hz,
and their analysis from 1990 to 2008 revealed a long-term change  in the multiplet  splittings  coinciding  with  the  1996 {\it sforzando} event,  where  the star  dramatically  altered its  pulsation  characteristics  on  a  timescale  of  hours.
These phenomena could be attributed to the interaction between 
convection and/or magnetic fields and pulsations. 

The {\it TESS} data of GD~358 presented in this work reveal the presence of 4 
of the 10 triplets found in \cite{1994ApJ...430..839W}. At variance with 
the results of that paper, in the {\it TESS} data we find
two complete triplets and two incomplete triplets. One of the complete triplets
has frequencies 2359.010~$\mu$Hz, 2362.689~$\mu$Hz, and 2366.318~$\mu$Hz. Going to smaller 
frequencies, we find the other complete triplet with frequencies 1623.248~$\mu$Hz, 1617.409~$\mu$Hz, and 1611.949~$\mu$Hz, an incomplete triplet with frequencies
1435.142~$\mu$Hz and 1421.059~$\mu$Hz\footnote{Between these two frequencies there is a frequency 
of 1426.986~$\mu$Hz but with a very low amplitude (1.294 ppt), which could be the $m= 0$ 
component of the triplet, although the splittings of the triplet would be quite different 
($\delta \nu \sim 8$~$\mu$Hz and $\delta \nu \sim 6$~$\mu$Hz). A peak of similar 
frequency (1427.27~$\mu$Hz) has been reported by \cite{1994ApJ...430..839W} with appreciable 
amplitude (19 ppt). However, we will not include the frequency of 1426.986~$\mu$Hz in this work 
since all periods detected in GD~358 from ground-based observations such as those of 
\cite{1994ApJ...430..839W} are taken into account through the "mean periods" calculated by 
\cite{2019ApJ...871...13B}; see Section \ref{pspacing-GD358}.}, and other 
incomplete triplet with frequencies
1297.338~$\mu$Hz and 1289.082~$\mu$Hz. These rotational triplets are emphasized with 
boldface in Table \ref{table:GD358-murat}.  We show the rotational triplets in 
Fig. \ref{fig:RM-GD358}. Similar to the findings of 
\cite{1994ApJ...430..839W}, we find that the frequency splittings $\delta \nu$ in 
these 4 triplets are not constant, as we will show in Sect. \ref{pspacing-GD358}.

\begin{table*}
\centering
\caption{Identified frequencies (combination frequencies), periods, and amplitudes (and their uncertainties)  and the 
signal-to-noise ratio in the  data of GD~358.}
\begin{tabular}{lcccr}
\hline
\noalign{\smallskip}
Peak & $\nu$    &  $\Pi$  &  $A$   &  S/N \\
 & ($\mu$Hz)      &  (s)   & (ppt)   &   \\
\noalign{\smallskip}
\hline
\noalign{\smallskip}

f$_{\rm 1}$    &  $919.507 \pm0.018$	&  $1087.538\pm0.021$	& $2.521\pm0.18 $	& 10.5 \\
f$_{\rm 2}$    &  $1038.177\pm0.026$	&  $963.226\pm0.025$	& $1.739\pm0.18 $	& 7.2  \\
f$_{\rm 3}$    &  $1082.770\pm0.008$	&  $923.556\pm0.007$	& $5.723\pm0.18 $	& 23.8 \\
f$_{\rm 4}$    &  $1170.146\pm0.015$	&  $854.593\pm0.011$	& $3.115\pm0.18 $	& 12.9 \\
f$_{\rm 5}$    &  $1232.928\pm0.004$	&  $811.076\pm0.003$	& $10.325\pm0.18$	& 43.0 \\
\hline
{\bf f$_{\rm 6}$}    &  $1289.082\pm0.006$	&  $775.745\pm0.003$	& $7.212\pm0.18 $	& 30.0 \\
{\bf f$_{\rm 7}$}    &  $1297.338\pm0.008$	&  $770.808\pm0.005$	& $5.574\pm0.18 $	& 23.2 \\
\hline
{\bf f$_{\rm 8}$}   &  $1421.059\pm0.004$	&  $703.700\pm0.002$	& $11.782\pm0.18$	& 49.0 \\
{\bf f$_{\rm 9}$}   &  $1435.142\pm0.018$	&  $696.794\pm0.009$	& $2.530\pm0.18 $	& 10.5 \\
\hline           
{\bf f$_{\rm 10}$}   &  $1611.949\pm0.010$	&  $620.367\pm0.004$	& $4.327\pm0.18 $	& 18.0 \\
{\bf f$_{\rm 11}$}   &  $1617.409\pm0.034$	&  $618.272\pm0.013$	& $1.352\pm0.18 $	& 5.6  \\
{\bf f$_{\rm 12}$}   &  $1623.248\pm0.013$	&  $616.048\pm0.005$	& $3.403\pm0.18 $	& 14.1 \\
\hline
f$_{\rm 13}$   &  $2024.182\pm0.020$	&  $494.026\pm0.005$	& $2.280\pm0.18 $	& 9.5  \\
f$_{\rm 14}$   &  $2154.064\pm0.010$	&  $464.238\pm0.002$	& $4.360\pm0.18 $	& 18.1 \\
f$_{\rm 15}$   &  $2157.584\pm0.042$	&  $463.481\pm0.009$	& $1.109\pm0.18 $	& 4.6  \\
\hline
{\bf f$_{\rm 16}$}   &  $2359.010\pm0.021$	&  $423.906\pm0.003$	& $2.214\pm0.18 $	& 9.2  \\
{\bf f$_{\rm 17}$}   &  $2362.689\pm0.021$	&  $423.246\pm0.003$	& $2.169\pm0.18 $	& 9.0  \\
{\bf f$_{\rm 18}$}   &  $2366.318\pm0.019$	&  $422.597\pm0.003$	& $2.380\pm0.18 $	& 9.9  \\
\hline
\hline
2f$_{\rm 1}$    &  $1839.301\pm0.031$	&  $543.684\pm0.009$	& $1.498\pm0.18 $	& 6.2  \\
f$_{\rm 3}$+f$_{\rm 5}$    &  $2315.669\pm0.036$	&  $431.840\pm0.006$	& $1.282\pm0.18 $	& 5.3  \\
2f$_{\rm 5}$    &  $2465.756\pm0.026$	&  $405.555\pm0.004$	& $1.772\pm0.18 $	& 7.3  \\
f$_{\rm 5}$+f$_{\rm 7}$    &  $2530.302\pm0.034$	&  $395.209\pm0.005$	& $1.337\pm0.18 $	& 5.5  \\
f$_{\rm 5}$+f$_{\rm 8}$   &  $2653.958\pm0.028$	&  $376.795\pm0.004$	& $1.614\pm0.18 $	& 6.7  \\
f$_{\rm 6}$+f$_{\rm 12}$ &          $2912.290\pm0.040$	&  $343.372\pm0.004$	& $1.147\pm0.18 $	& 4.7  \\
f$_{\rm 8}$+f$_{\rm 12}$  &  $3044.295\pm0.029$	&  $328.483\pm0.003$	& $1.586\pm0.18 $	& 6.6  \\
f$_{\rm 5}$+f$_{\rm 14}$   &  $3386.984\pm0.038$	&  $295.247\pm0.003$	& $1.213\pm0.18 $	& 5.0  \\

\noalign{\smallskip}
\hline
\end{tabular}
\label{table:GD358-murat}
\end{table*}

\section{Evolutionary models, numerical codes, and spectroscopic masses}
\label{models}

We employ a  set of fully evolutionary DB WD stellar  models that consider
the  whole evolution of the progenitor stars. Specifically, the stellar 
models were taken  from the evolutionary  calculations presented by  
\cite{2009ApJ...704.1605A} produced with the 
{\tt LPCODE} evolutionary code. For details about the input physics
and evolutionary code, and  the   numerical  simulations  performed
to   obtain  the  DB WD evolutionary sequences  employed here, we
refer  the interested reader to that paper. These evolutionary 
tracks have been employed in the asteroseismic analyses of the DBV stars KIC\,8626021 
\citep{2012A&A...541A..42C}, KUV 05134+2605 \citep{2014A&A...570A.116B},  
PG\,1351+489 \citep{2014JCAP...08..054C}, and  WD 0158$-$160 \citep{2019A&A...632A..42B}. 
The sequences of DB WD models have been computed considering a detailed 
treatment of  the prior evolution, starting from the 
zero-age main sequence (ZAMS), taking into account the thermally pulsing 
asymptotic giant branch (TP-AGB) and born-again (VLTP; very late thermal pulse) 
phases, the stage of the PG\,1159 stars, and finally the DB WD phase. 
By virtue of this, the models have evolving chemical profiles consistent 
with the prior evolution. The models assume the ML2 prescription of convection with the mixing length parameter, $\alpha$, fixed to 1 \citep{1971A&A....12...21B,1990ApJS...72..335T}.
Specifically, we considered nine DB WD sequences with stellar
masses of:  $0.515, 0.530, 0.542, 0.565, 0.584,  0.609, 0.664, 0.741$,
and $0.870  M_{\odot}$.  These DB WD sequences are characterized  by the
maximum He-rich  envelope that  can be left  by prior evolution  if we
assume that they are the result  of a born-again episode. The value of
envelope  mass ranges  from  $M_{\rm He}/M_*  \sim  2 \times  10^{-2}$
($M_*=  0.515 M_{\odot}$) to  $M_{\rm He}/M_*  \sim 1  \times 10^{-3}$
($M_*=  0.870  M_{\odot}$).   In Figure \ref{fig:1} we 
show the complete set of DB WD evolutionary tracks (with different
colors according to the value  of the stellar  mass) along with the 
location  of  all  the  DBVs known  to  date \citep{2019A&ARv..27....7C}, 
including GD~358.    

The $\ell=  1, 2$ $g$-mode  pulsation  periods employed in this work
were computed with the  adiabatic and nonadiabatic versions of 
the pulsation code {\tt LP-PUL}
\citep[][]{2006A&A...454..863C,2006A&A...458..259C,2009JPhCS.172a2075C} 
and the same
methods  employed  in the  previous  works of de La Plata Stellar
Evolution and Pulsation Research Group\footnote{\tt
http://fcaglp.fcaglp.unlp.edu.ar/evolgroup/}.  

We assess a value of the spectroscopic mass of GD~358 by interpolation 
on the   evolutionary    tracks   presented   in
Fig.~\ref{fig:1} and the  published values of the spectroscopic 
surface gravity and temperature. This is a relevant aspect 
because  this same  set  of  DB  WD  models  is  employed  
to derive the stellar mass from the period spacing (next Section).  
We get an spectroscopic stellar mass of $M_{\star}= 0.508 \pm 0.050
\ M_{\sun}$ if $T_{\rm eff}= 24\,000\pm 500$~K and $\log g= 7.80\pm 0.05$ 
\citep{2012ASPC..462..171N}, and $M_{\star}= 0.560 \pm 0.028\ M_{\sun}$ if $T_{\rm eff}= 24\,937\pm 1\,018$~K 
and $\log g= 7.92\pm 0.05$ \citep{2017ApJ...848...11B}.    
The uncertainties in the stellar mass are derived from the 
errors in $T_{\rm eff}$ and $\log g$ adopting the extreme values of 
these parameters when interpolating between the evolutionary tracks 
of Fig. \ref{fig:1}.

\section{Asteroseismology} 
\label{asteroseismology}

In the asymptotic limit of high-radial orders ($k \gg \ell$), the periods of 
$g$ modes  with consecutive radial order are approximately evenly separated 
\citep{1990ApJS...72..335T}, being the constant period spacing dependent on the 
harmonic degree:

\begin{equation}\label{eq:1}
\Delta \Pi_{\ell}^{\rm a} = \frac{{\Pi}_{0}}{\sqrt{\ell(\ell+1)}},
\end{equation} 

\noindent $\Pi_{0}$ being a constant value defined as:

\begin{equation}\label{eq:2}
\Pi_{0}=  \frac{2 \pi^2}{\left[ \int_{r_1}^{r_2}
\frac{N}{r} dr \right]}, 
\end{equation} 

\noindent where $N$ is the Brunt-V\"ais\"al\"a frequency. 
The asymptotic period spacing given by Eq. (\ref{eq:1}) is very close 
to the computed period spacing  of $g$ modes in chemically homogeneous 
stellar models without convective regions
\citep{1980ApJS...43..469T}.  In the case of pulsating DB WDs, they  may have 
a surface convective zone, although usually very thin 
compared to the stellar radius. On the other hand, they have several 
chemical composition gradients. Mainly for this last reason, the calculated 
period spacing does not coincide with the asymptotic period spacing given by 
Eq. (\ref{eq:1}), but nevertheless the {\it average} of the calculated 
spacing is very close to $\Delta \Pi_{\ell}^{\rm a}$ for radial orders high 
enough. The departures of the 
period spacing from the averaged period spacing are provoked by the 
mechanical resonance  called "mode trapping".  
Mode trapping  has been intensively studied in the
context of  DAV, DBV, and GW Vir  stars \citep[see,
e.g.,][]{1992ApJS...80..369B, 1993ApJ...406..661B, 1994ApJ...427..415K, 
2002A&A...387..531C, 2006A&A...454..863C}.  

The methods we use in this paper to extract information of the stellar
mass and the internal structure of GD~358 
are the same employed in  \cite{2021A&A...645A.117C} for GW Vir stars  observed with {\it TESS} \citep[see, also,][] {2012A&A...541A..42C,2014A&A...570A.116B,
2014JCAP...08..054C, 2019A&A...632A..42B}. Specifically, 
we compare the observed period spacing of GD~358 ($\Delta \Pi$) 
with the asymptotic period spacing ($\Delta \Pi_{\ell}^{\rm a}$) computed with
Eq.~(\ref{eq:1}) at the effective temperature of the star to derive
an estimate of the stellar mass. DBV stars generally do 
not have all of their pulsation modes in the asymptotic regime, so there 
is usually no perfect agreement between $\Delta \Pi$ and 
$\Delta \Pi_{\ell}^{\rm a}$.   Therefore, the
derivation of the  stellar  mass  using  the  asymptotic period
spacing  may  not be entirely reliable in DBV stars that
pulsate with modes characterized by low and intermediate radial
orders, but it gives a good estimate of the stellar mass for stars
pulsating with $g$ modes of high radial order \citep[see][for the case of 
GW Vir stars]{2008A&A...478..175A}. A variation of this approach to
infer the stellar mass of DBV stars is to compare $\Delta \Pi$ with
the average of the computed period spacings ($\overline{\Delta
\Pi_{k}}$). It  is calculated  as $\overline{\Delta \Pi_{k}}= (n-1)^{-1} \sum_k \Delta \Pi_{k}$, where the "forward" period spacing ($\Delta \Pi_{k}$) is
defined as $\Delta \Pi_{k}= \Pi_{k+1}-\Pi_{k}$ ($k$ being the radial
order) and $n$ is the number of computed periods laying in the range
of the observed periods.  This method is more reliable for the
estimation of the stellar mass of DBV stars than that described
above using $\Delta \Pi_{\ell}^{\rm a}$ because, provided that  the
average of  the  computed period  spacings is evaluated at the
appropriate range of periods, the approach is valid for the regimes of
short, intermediate and long periods as well. When the average of the
computed period spacings is taken over a range of periods
characterized by high $k$ values, then the predictions of the present
method become closer to those of the asymptotic period-spacing
approach \citep[][]{2008A&A...478..175A}.  On the other hand, the
present method requires detailed period computations, at variance
with the method described above, which does  not  involve  pulsational
calculations.  Note that both methods for assessing the stellar mass
rely on the spectroscopic effective temperature, and the results are
unavoidably affected by its associated uncertainty. The methods 
outlined above take full advantage of the fact that the period spacing of 
DBV stars primarily depends on the stellar mass and the effective temperature, 
and very weakly on the thickness of the He envelope 
\citep[see, e.g.,][]{1990ApJS...72..335T}\footnote{These methods cannot, in principle,
be directly applied to DAV stars to infer the stellar mass, for which 
the period spacing depends simultaneously on $M_{\star}$, $T_{\rm eff}$ and $M_{\rm H}$ with comparable sensitivity, and this implies the existence of  multiple combinations of these three quantities that produce the same spacing of periods.}. 

A powerful approach to study the internal structure
of pulsating stars is to search for models that best fit the
observed pulsation  periods. To quantify  the
goodness of the match between the theoretical periods
($\Pi_{\ell,k}$) and the observed periods ($\Pi_i^{\rm
  o}$), we follow the same procedure as in our previous studies: 

\begin{equation}
  \chi^2(M_{\star}, T_{\rm eff})= \frac{1}{N} \sum_{i= 1}^{N}
      {\rm min}[(\Pi_{\ell,k}-\Pi_i^{\rm o})^2]
\label{eq:3}
\end{equation}  

\noindent $N$ being the number of observed periods. In order to find the
stellar model that best fits the observed periods exhibited by
GD~358 --the ``asteroseismological'' model-- we
evaluate the  function  $\chi^2$ for stellar  masses  $M_{\star}=
0.515, 0.530, 0.542,  0.565, 0.584, 0.609, 0.664$, $0.741, 0.870 M_{\odot}$. For
the  effective  temperature, we  employ  a very fine model grid ($\Delta
T_{\rm eff}= 10-30$ K).  The DB WD model that
shows the smallest value of $\chi^2$ is adopted as the best-fit
asteroseismological model. Below, we employ the tools described above to extract 
information of GD~358. 

\subsection{Period spacing and the seismic mass}
\label{pspacing-GD358}

Ground-based photometric observations of GD~358 span a period of 34 years. 
No other DBV star has been studied for such a long period of time, which is why we know the 
most about this object. A detailed compilation  of the observations, that includes archival data from 1982 to 2006, and 
1195.2 hr of new observations from 2007 to 2016, 
has been presented by \cite{2019ApJ...871...13B}. 
Figure 4 of that paper, constructed 
on the basis of the periods and amplitudes of their Tables 2 and 3, is extremely illustrative of how the periods of pulsation in this star are concentrated in bands with finite widths, rather than discrete periods. Frequencies detected in a given observing season are not found in all observing runs, and most of the detected frequencies are not statistically identical from year to year. Only a few frequencies appear with exactly the same values at various observing runs.
Regarding the bands of periods exhibited by GD~358, 
\cite{2019ApJ...871...13B} find a general increase in their width with decreasing frequency (increasing period), at least until the band at $1238~\mu$Hz (807 s). 
This behaviour could be related to the oscillation of the outer convection zone of the WD during pulsations \citep{2020ApJ...890...11M}. Specifically, the oscillation 
in the base of the convection zone would affect the radial eigenfunction of 
$g$ modes that have the outer turning point of oscillation located precisely at the 
base of the outer convection zone. We note, however, that
if the reason for the existence of bands of periods in GD~358 is
the oscillation of the base of the outer convective zone, then the
{\it overall structure} of the star remains largely unchanged.

\begin{figure} 
\includegraphics[clip,width=1\columnwidth]{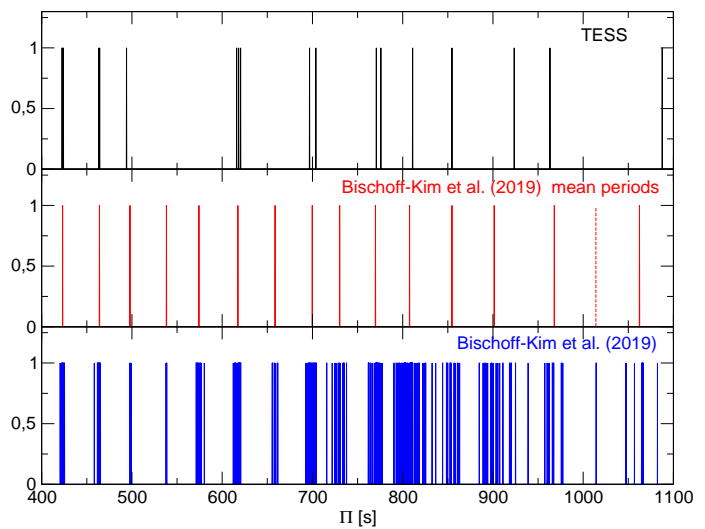}
\caption{Illustrative distribution of the periods of GD~358
  according  to {\it TESS} (19 periods, black lines, upper panel), and
  according to \cite{2019ApJ...871...13B}  (15 mean periods, red lines,
  middle panel, and 352 periods, blue lines, lower panel). The period
  at 1014.35 s in the middle panel, represented with a dashed red line, 
  corresponds to a single detection in the 2016 ground-based observations.
  The amplitudes have been arbitrarily  set to one for
  clarity.}
\label{fig:compara-tess-bk19} 
\end{figure}

While the origin of this phenomenon is not entirely clear and deserves further exploration, we note that the existence of finite bands of periods poses a problem for applying the asteroseismological tools, because they require a set of {\it discrete} observed periods (even with possible uncertainties) that must be compared with well-defined periods calculated in stellar models of WDs. To tackle this difficulty,  \cite{2019ApJ...871...13B}  determined the mean periods for each band to be used in the asteroseismic fits (see their Table 5).  The use of the mean values of periods in asteroseismology is well-justified given that the periods do not change secularly, but remain within a limited range of periods forming each band.
In the lower and middle panels of Fig. \ref{fig:compara-tess-bk19}, we show 
schematically all the periods detected from ground-based observations and 
the mean periods determined by \cite{2019ApJ...871...13B} for each band, respectively. 
The amplitude has been arbitrarily set to one to facilitate
visualization. The period at 1014.35 s in the middle panel, 
represented with a dashed red line, is not associated to any specific band,
but instead corresponds to a single detection in the 2016 
ground-based observations. In the upper panel of Fig. \ref{fig:compara-tess-bk19} 
we include the 18 periods detected by {\it  TESS} (Table \ref{table:GD358-murat}).
By comparing the mean periods (middle panel) and the periods detected by {\it TESS} 
(upper panel), we note that  there are at least 9 periods that coincide 
between both sets, at $\sim 420$ s, $\sim 465$ s, $\sim 495$ s, $\sim 620$ s, $\sim 700$ s, $\sim 770$ s, $\sim 810$ s, $\sim 855$ s, and $\sim 970$ s.
Other periodicities are present in the ground-based mean periods but not in the 
{\it TESS} periods ($\sim 540$ s, $\sim 575$ s, $\sim 660$ s, $\sim 730$ s, $\sim 902$ s, $\sim 1015$ s, and $\sim 1063$ s), and vice versa ($\sim 925$ s and $\sim 1090$ s). 
The space-based detection of 9 periods already found from ground-based observations   
is a great finding itself, because it confirms the results derived from exhaustive previous studies. Likewise, the detection of 2 additional signals which were not detected in previous works, allows us to broaden the spectrum of periods available for the asteroseismological study of this star. In fact, in order to extract as much information as possible with the tools of
asteroseismology, it is crucial to employ as many periods (which represent eigenvalues of the star) as possible. In order to identify the pulsation 
modes and determine the possible period spacing of GD~358, which is essential to estimate the stellar mass, we consider
an expanded list of periods by combining the 15 dipole $m=0$ mean periods found by
\cite{2019ApJ...871...13B}  to the set of 
periods measured by  {\it TESS} (Table \ref{table:GD358-murat}).  
In the case of the 9 periods close to each other detected in both data sets, 
we decided to adopt the  
  periods measured by {\it TESS} because  they are in
general more accurate. In the case of the periods near $700$~s, we adopt 
the period 699.82~s from \cite{2019ApJ...871...13B}, which seems to be the central 
component of the incomplete rotational triplet $(+1,0,-1)=(696.794\ {\rm s},..., 703.700\ {\rm s})$ 
detected by {\it TESS} (see below).   The
resulting extended list of periods to be used in our analysis contains 26
periods and is shown in the first and second columns of 
Table \ref{table:GD358-extended}. Note that we have 
also considered the period 1014.35 s, which has been detected only in the 
observations of 2016. This period was not considered in the asteroseismological 
analysis of \cite{2019ApJ...871...13B}. However, since its value seems to fit very 
well in the apparent pattern of dipole periods with constant separation 
present in this star, as can be guessed from Fig. \ref{fig:compara-tess-bk19},  
we decided to include it in our subsequent analysis.  We consider 
a proper procedure to combine periods detected at different
times and both with observations from the ground and from space. We
have the excellent example of the DAV star G29$-$38 \citep{1998ApJ...495..424K} 
where a global pulsation spectrum for this star was constructed using 
different ground-based observations from a decade. Whatever the physical 
mechanism behind the alternate appearance and disappearance of modes, 
we can perform robust asteroseismological analyses by collecting  
the data from all epochs and constructing a combined spectrum of pulsations.
This has been demonstrated for the DBV star KIC~08626021 
\citep[see][]{2011ApJ...742L..16B,2014ApJ...794...39B,2018Natur.554...73G}.

In the case of the 
rotational triplets, we have assigned the $m$  value to the different components 
following \cite{1994ApJ...430..839W}. We note
that the frequency splittings are not constant among the different triplets. In fact,  
we have $\delta \nu= 3.679\ \mu$Hz and $\delta \nu= 3.629\ \mu$Hz for the complete 
triplet centered at $2362.689\ \mu$Hz (423.246 s), 
$\delta \nu= 5.839\ \mu$Hz and $\delta \nu= 5.460\ \mu$Hz for the complete triplet centered at 
$1617.409\ \mu$Hz (618.272 s), $2\ \delta \nu= 14.083 \ \mu$Hz (that is, 
$\delta \nu= 7.042 \ \mu$Hz) 
for the incomplete triplet with side components  
$1435.142\ \mu$Hz (696.794 s) and $1421.059\ \mu$Hz (703.700 s), and 
$\delta \nu= 8.256\ \mu$Hz for the incomplete triplet centered at $1289.082\ \mu$Hz (775.745 s). 
There is an apparent trend of larger 
$\delta \nu$ for decreasing frequencies (increasing periods), in agreement
with \cite{1994ApJ...430..839W} (see their Fig. 6 and Table 2). This dependence of the frequency splittings with the radial order of the modes could indicate differential rotation of GD~358, 
since each mode samples areas of different depth in the star, and would 
indicate different speeds of rotation\footnote{We note that it makes no sense to 
calculate different rotational periods for GD~358 using the different frequency splittings using an  uniform-rotation formula like the one described in Sect. \ref{observations}.}.
While these results would imply that GD~358 does not rotate as a 
rigid body, in order to put this results on a firm basis it would be necessary to make a detailed analysis such as that carried out by \cite{1999ApJ...516..349K} 
\citep[see, also,][for the specific case of the GW Vir star 
PG~122+200]{2011MNRAS.418.2519C}, 
which is beyond the scope of this paper. 

\begin{table}
\centering
\caption{Enlarged list of periods of GD~358. Column 1 corresponds to
  seven $\ell= 1$ $m= 0$ average periods derived by \cite{2019ApJ...871...13B}
  (BK19). We include also the period at 1014.35 s extracted 
  from the ground based observations of 2016 \citep[see Table 3 of ][]{2019ApJ...871...13B}.
  Column 2 corresponds to  18 periods detected by {\it TESS}   
  (Table \ref{table:GD358-murat}).  The periods with an asterisk are the 13 
  periods used in the linear least square fit depicted in Fig. \ref{fig:fit-GD358}.}
\begin{tabular}{cc|cccr}
\hline
\noalign{\smallskip}
$\Pi_i^{\rm O}$ (s) & $\Pi_{i}^{\rm O}$ (s) & $\Pi_{\rm fit}$ (s) & 
$\delta\Pi$ (s) & $\ell^{\rm O}$ & $m^{\rm O}$\\
  BK19 & {\it TESS} & & & & \\
\noalign{\smallskip}
\hline
\noalign{\smallskip}       
         & 422.597   &         &          & 1 & $+1$ \\ 
         & 423.246*  & 421.697 &  1.549   & 1 &  0 \\
         & 423.906   &         &          & 1 & $-1$ \\
         & 463.481*  & 460.942 &  2.539   & 1 &  0 \\
         & 464.238   &         &          & ? &  ? \\
         & 494.026   &         &          & ? &  ? \\
538.30*  &           & 539.433 &  $-1.133$  & 1 &  0 \\
574.22*  &           & 578.678 &  $-4.458$ & 1 &  0 \\
         & 616.048   &         &          & 1 & $+1$ \\
         & 618.272*  & 617.923 &  0.349   & 1 &  0 \\
         & 620.367   &         &          & 1 & $-1$ \\
658.69*  &           & 657.168 &  1.522   & 1 &  0 \\
         & 696.794   &         &          & 1 & $+1$ \\
699.82*  &           & 696.414 &  3.406   & 1 & 0  \\
         & 703.700   &         &          & 1 & $-1$ \\
730.28*  &           & 735.659 &  $-5.379$  & 1 &  0 \\
         & 770.808   &         &          & 1 & $+1$ \\
         & 775.745*  & 774.904 &  0.841   & 1 &  0 \\
         & 811.076*  & 814.149 &  $-3.073$  & 1 &  0 \\
         & 854.593*  & 853.394 &   1.199  & 1 &  0 \\
901.49   &           &         &          & ? &  ? \\ 
         & 923.556   &         &          & ? &  ? \\
         & 963.226   &         &          & ? &  ? \\   
1014.35*  &          & 1010.575  & 3.775  & 1 &  0 \\    
1062.32   &          &           &         & ? &  ? \\
          & 1087.538* & 1088.866 &  $-1.328$ & 1 & 0 \\
\noalign{\smallskip}
\hline
\end{tabular}
\label{table:GD358-extended}
\end{table}

\begin{figure} 
\includegraphics[clip,width=1.0\columnwidth]{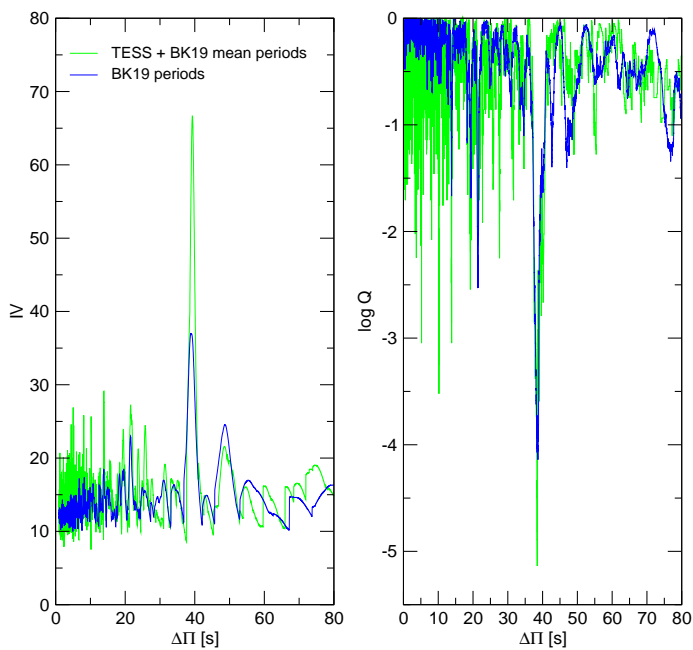}
\caption{ I-V  (left panel) and  K-S  (right panel) 
significance  tests  to  search  for  a constant  period
  spacing  in GD~358. 
  The tests are applied to the complete set of 352 pulsation periods of 
  Tables 2 and 3 of \cite{2019ApJ...871...13B}
(blue curves), and to the combination of the 
{\it TESS} periods plus the mean periods of \cite{2019ApJ...871...13B} (green curves), 
included in Table \ref{table:GD358-extended}. A clear signal of a constant period spacing at  $\sim 39$~s is 
  evident. See text for details.}
\label{fig:tests-GD358} 
\end{figure}

\begin{figure} 
\includegraphics[clip,width=1.0\columnwidth]{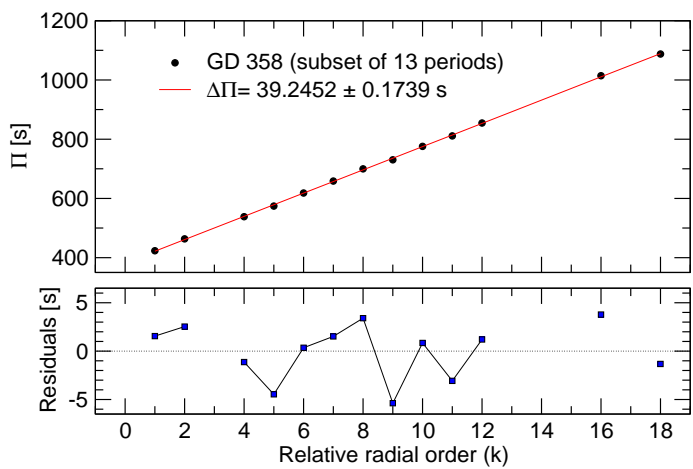}
\caption{Upper panel: linear least-squares fit to the 13 periods of
  GD~358 marked with asterisks in Table
  \ref{table:GD358-extended}. The derived period spacing from this
  fit is $\Delta \Pi= 39.25\pm 0.17$~s.  Lower panel: the residuals
  of the period distribution relative to the mean period spacing,
  revealing signals of mode trapping in the period spectrum of
  GD~358. Modes with consecutive radial order are connected with 
  thin black lines.}
\label{fig:fit-GD358} 
\end{figure} 

We  looked for a  constant  period  spacing  in  the  data of
GD~358  using the  Kolmogorov-Smirnov
\citep[K-S;][]{1988IAUS..123..329K}, and the inverse  variance
\citep[I-V;][]{1994MNRAS.270..222O}  statistical tests.  
Fig.~\ref{fig:tests-GD358} displays the results of applying the K-S
and I-V significance tests to the period spectrum of GD~358. We show 
the case in which we adopted the full set of 352 periods of Tables 2 and 3 of \cite{2019ApJ...871...13B} (blue curves), and the situation in which 
the {\it TESS} periods plus the mean periods 
of \cite{2019ApJ...871...13B} (green curves) are considered.
In this last case we also considered the period at 1014.35 s disregarded in that paper.
The two tests 
point to the  existence of a pattern of $\ell= 1$ constant period spacing of $\Delta \Pi\sim
39$~s.

To derive a refined value of the period  spacing, we have
carried out a linear least-squares fit to the 13 periods marked  with
an asterisk in Table \ref{table:GD358-extended} (see Fig. \ref{fig:fit-GD358}). 
These periods are all supposed to be the $\ell= 1$ $m= 0$ members of a 
sequence of periods equally spaced. 
We obtain a period spacing  of $\Delta \Pi= 39.25\pm 0.17$~s.  
This period spacing corresponds to our expectations for a dipole ($\ell= 1$) 
sequence. On the other hand, if we assume that the 
$\sim 39$~s period spacing 
were due to quadrupole modes, then the dipole period spacing would be 
$\sim 68$~s, which is not present in the tests\footnote{If such a dipole period spacing 
of $\sim 68$~s existed, it would involve an extremely low stellar mass for GD~358, 
which can safely be ruled out.}. Thus, the identification of the $\sim 39$~s period 
spacing as due to a sequence of dipole modes is robust. This sequence includes the 
13 periods marked with an asterisk in Table \ref{table:GD358-extended}, and 
the 7 periods that are the $m= -1$ or $m= +1$ components of the four triplets. 
The remaining 6 periods in that table can be associated with $\ell= 1$
modes which, due to mode trapping effects, deviate substantially from the derived
sequence of almost equally-spaced periods, or with modes with $\ell=
2$ (or possibly higher) modes. In short, we have a total of 20 periods
identified with $\ell= 1$ modes. 

In the lower panel of Fig. \ref{fig:fit-GD358} we show the residuals
($\delta \Pi$) between the dipole observed periods ($\Pi_i^{\rm O}$) and the
periods derived from the mean period spacing ($\Pi_{\rm fit}$). 
The average of the
absolute values of the residuals is $\overline{|\delta \Pi|}= 2.35$ s. 
The presence of  several minima in the distribution of residuals strongly
suggests the mode-trapping  effects inflicted by the presence of
internal chemical transition regions.

\begin{figure} 
\includegraphics[clip,width=1.0\columnwidth]{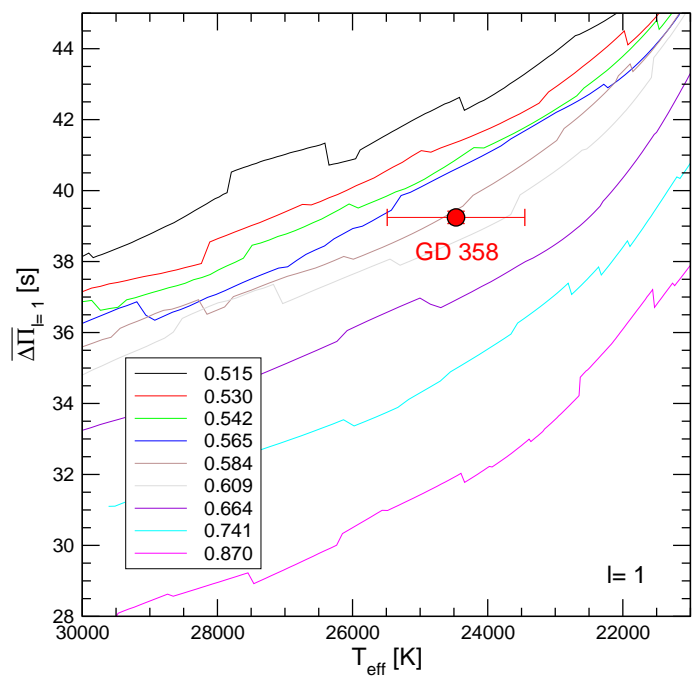}
\caption{Dipole ($\ell= 1$) average of the computed period spacings,
  $\overline{\Delta \Pi_k}$, evaluated  in  a  range  of  periods  that
  embraces  the  periods  observed  in GD~358, shown as curves 
  of different colors for different stellar masses. 
  We consider a mean effective temperature for the star, 
  $T_{\rm eff}= 24\,469\pm 1018$~K, resulting from averaging 
  $T_{\rm eff}= 24\,000\pm 500$~K \citep{2014A&A...568A.118K} 
  and $T_{\rm eff}= 24\,937\pm 1\,018$~K \citep{2017ApJ...848...11B}. 
  We adopt  the mean period spacing   $\Delta \Pi= 39.25\pm 0.17$~s derived in
  Sect.~\ref{pspacing-GD358}. 
  We include the error bars associated to the uncertainties in
  $\overline{\Delta \Pi_k}$ and $T_{\rm eff}$.  The stellar mass
  derived by interpolation is $M_{\star}= 0.588\pm0.024 M_{\odot}$.}
\label{fig:psp-teff-GD358} 
\end{figure}

\begin{figure} 
\includegraphics[clip,width= 1.0\columnwidth]{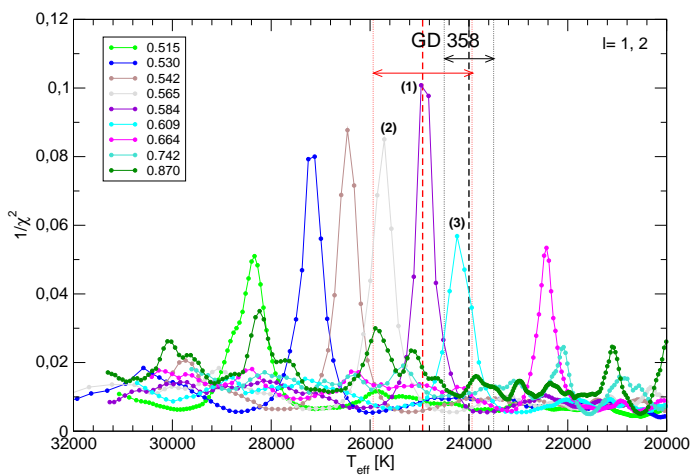}
\caption{The inverse of the quality function of the period fit in terms of 
the effective temperature, shown with different colors for the different 
stellar masses. The vertical black dashed line corresponds to 
the spectroscopic $T_{\rm eff}$ of GD~358 and the vertical dotted lines 
its uncertainties \citep[$T_{\rm eff}= 24\,000\pm 500$ K;][]{2014A&A...568A.118K}. 
Analogously, the blue vertical lines correspond to the spectroscopic $T_{\rm eff}$ 
and its uncertainties as given by \cite{2017ApJ...848...11B} 
($T_{\rm eff}= 24\,937\pm 1\,018$ K). 
Three maxima have been labeled as (1), (2) and (3), corresponding 
to the three asteroseismological solutions compatible with spectroscopy 
(see the text).}
\label{fig:chi2-GD358} 
\end{figure} 

We calculated the average of the computed period spacings for
$\ell= 1$, $\overline{\Delta \Pi_{k}}$,  in terms of the effective
temperature for all the masses considered and a period interval of $400-1100$ s,
corresponding to the range of periods exhibited by GD~358. 
The results  are shown in Fig.~\ref{fig:psp-teff-GD358}, where we depict  $\overline{\Delta
\Pi_{k}}$ with curves of different colors according to the various stellar 
masses. For the location of GD~358, indicated by a small red  circle with error 
bars, we considered the average of the effective temperature, 
$T_{\rm eff}= 24\,469\pm 1018$ K, based on 
the effective temperatures derived by \cite{2017ApJ...848...11B} and 
\cite{2014A&A...568A.118K}. We perform a linear interpolation and 
obtain $M_{\star}= 0.588 \pm 0.024\,M_\odot$.

\subsection{Period fits and the asteroseismological model}
\label{p-to-p-fits-GD358}

In our analysis of period-to-period fits, we only
took into account the components $m= 0$ present in the frequency spectrum of GD~358,  
and ignored the $m\neq 0$ components. In total, we have 19 observed periods as 
the input for our asteroseismological period fits (see column 1 of Table 
\ref{table:GD358-asteroseismic-model}). According to the results derived 
in Sect.~\ref{pspacing-GD358}, we can assume that a subset of 13 $m= 0$ periods 
are associated with $\ell= 1$ modes --- those marked with an asterisk in
Table~\ref{table:GD358-extended}--- and leave free the identification  
as $\ell = 1$ or $\ell = 2$ for the remaining 6 periods. We display our results in
Fig.~\ref{fig:chi2-GD358}. We find three possible solutions, that is,  
maxima of $(\chi^2)^{-1}$, labelled as (1), (2), and (3),  
which are compatible with both effective temperature determinations of GD~358 and its
uncertainties. These solutions are (1)  $T_{\rm eff}= 24\,967$ K 
and $M_{\star}= 0.584\ M_{\odot}$ 
($\chi^2= 9.921$),  (2)  $T_{\rm eff}= 25\,712$ K and $M_{\star}= 0.565\ M_{\odot}$ 
($\chi^2= 11.758$), and (3)  $T_{\rm eff}= 24\,240$ K and $M_{\star}= 0.609\ M_{\odot}$ 
($\chi^2= 17.595$). Clearly, the optimal solution is (1), since the DB WD
model associated to it provides the best agreement between the 
theoretical and observed periods. Note that the effective temperature of this model 
is very close to that of GD~358 according to the spectroscopic determination of
\cite{2017ApJ...848...11B}. 

We adopt the model characterized by $M_{\star}= 0.584 M_{\odot}$,
$T_{\rm eff}= 24\,967$~K,  and $\log(L_{\star}/L_{\odot})= -1.215$ as
the asteroseismological model for GD~358.  The location of this model in the $\log g- T_{\rm eff}$ diagram is displayed in Fig. \ref{fig:1} with a blue circle. In
Table~\ref{table:GD358-asteroseismic-model} we show a  detailed
comparison of the observed periods of GD~358 and the theoretical
$m= 0$ periods  of  the  asteroseismological  model. According to this 
model, the  periods exhibited by
the star correspond to 16 dipole $m= 0$ modes with radial order $k$ in the range
$k \in [8,25]$, and 3 quadrupole modes with $36 \leq k \leq 43$. The average of 
the computed $\ell= 1$ period spacings  for this model is 
$\overline{\Delta \Pi}_{\ell= 1}= 38.926$ s, very similar to the dipole 
mean period spacing obtained in Sect. \ref{pspacing-GD358} for
this star, $\Delta \Pi= 39.25 \pm 0.17$ s. In order to
quantitatively assess the goodness of our period-to-period fit, we compute the
average   of   the   absolute   period    differences,
$\overline{\delta \Pi_i}= \left( \sum_{i= 1}^n |\delta \Pi_i|
\right)/n$, where $\delta \Pi_i= (\Pi_{\ell,k}-\Pi_i^{\rm o})$ and $n=
19$,  and the root-mean-square residual, $\sigma= \sqrt{(\sum_{i= 1}^n
  |\delta \Pi_i|^2)/n}= \sqrt{\chi^2}$.  We obtain $\overline{\delta
  \Pi_i}= 2.56$~s and $\sigma= 3.15$~s.  To have a global indicator of the
quality of the period fit that takes into account the number of free
parameters, the number of fitted periods, and the proximity between
the  theoretical and observed periods, we computed the Bayes
Information Criterion \citep[BIC;][]{2000MNRAS.311..636K}\footnote{See \cite{2004MNRAS.351L..49L,2007MNRAS.377L..74L} for an equivalent 
formulation of the BIC index.}: 

\begin{equation}
{\rm BIC}= n_{\rm p} \left(\frac{\log N}{N} \right) + \log \sigma^2,
\end{equation}

\noindent $n_{\rm p}$ being the number of free parameters of the
models, and $N$ the number of observed periods. The smaller the
value of BIC, the better the quality of the fit.  Note that this 
criterion penalizes for an excess of parameters. In our case, $n_{\rm
  p}= 2$ (stellar mass and effective temperature),   $N= 19$, and
$\sigma= 3.15$\,s.   We obtain ${\rm BIC}= 1.13$ for the 
asteroseismological model, this value being the smallest among the 
possible solutions (1), (2) and (3) shown in Fig. \ref{fig:chi2-GD358}. 
Also, the obtained BIC value is similar to that derived by \cite{2019ApJ...871...13B} 
(BIC= 1.2) for their best period fit to GD~358. The low value of BIC obtained 
in this work indicates that  our period fit is very good.

\begin{figure} 
\includegraphics[clip,width= 1.0\columnwidth]{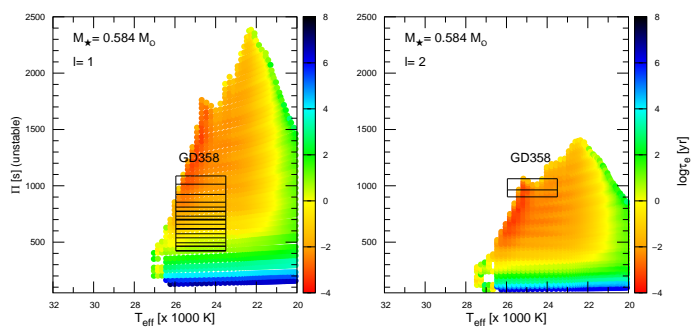}
\caption{Left panel: periods of excited $\ell= 1$ $g$ modes as a function of the effective 
temperature, with the palette of colors (right scale) indicating the 
logarithm of the $e$-folding time ($\tau_{\rm e}$ in years), for  
the DB WD sequence with $M_{\star}= 0.584 M_{\odot}$. Right panel: same as left panel, 
but for $\ell= 2$ modes. In both panels, the pulsation periods of
the DBV star GD~358 with the identification of $\ell$ according to 
our asteroseismological model (see Table \ref{table:GD358-asteroseismic-model}), 
are shown as horizontal segments, where their widths represent 
the possible $T_{\rm eff}$ interval, according to spectroscopy.}
\label{fig:per-teff-eft-0584-GD358} 
\end{figure}

Table~\ref{table:GD358-asteroseismic-model} also includes the
secular rates of period change ($\dot{\Pi}\equiv d\Pi/dt$) expected for each
$g$ mode of GD~358. Note that all of them are positive
($\dot{\Pi}>0$), meaning that the periods are lengthening over
time. The rate of  change of periods in WDs and pre-WDs is related
to $\dot{T}$ ($T$ being the temperature at the region of the period
formation) and $\dot{R_{\star}}$ ($R_{\star}$ being the stellar
radius) through the approximate expression $(\dot{\Pi}/\Pi) \approx -a\ (\dot{T}/T) + b\ (\dot{R_{\star}}/R_{\star})$
\citep{1983Natur.303..781W}. According  to our asteroseismological
model, the star is cooling with  approximately constant stellar radius. 
As a consequence, $\dot{T}<0$ and $\dot{R_{\star}} \sim 0$, and then,
$\dot{\Pi} > 0$. Can we compare these theoretical estimates of the 
secular drift of the periods with the true rates of period change of GD~358? 
The frequencies (periods) $2362.689\ \mu$Hz (423.246 s) and 
$2154.064\ \mu$Hz (464.238 s) of GD~358, 
corresponding to the $g$ modes with $k= 8$ and $k= 9$ according to our 
asteroseismological model,  are the most stable frequencies of this star 
\citep{2003A&A...401..639K, 2009ApJ...693..564P}. 
However, the frequency shifts are large enough to mask any possible signs 
of evolutionary period change. Thus, we are forced to conclude that 
the observed periods of GD~358 are not stable enough
to be able to measure the rate of change of periods 
due to the evolution of the star. This means that the values derived for 
the rate of change of periods of our seismological model of GD~358 remain 
(for now) only of academic interest.

\begin{table*}
\centering
\caption{Observed and theoretical $m= 0$ periods of the asteroseismological
  model for GD~358 [$M_{\star}= 0.584 M_{\odot}$, $T_{\rm eff}=
     24\,967$ K, $\log(L_{\star}/L_{\odot})= -1.215$] corresponding to 
     solution (1) in Fig. \ref{fig:chi2-GD358}. 
     $\delta \Pi_k= \Pi^{\rm O}_i-\Pi_k$ represents  the
  period differences, $\ell$ the harmonic degree, and $k$ the radial
  order.  The last column gives information
  about the pulsational stability/instability   nature  of  the
  modes.}
\begin{tabular}{cc|cccrcc}
\hline
\noalign{\smallskip}
$\Pi_i^{\rm O}$ & $\ell^{\rm O}$ & $\Pi_k$ & $\ell$ & $k$ &  $\delta \Pi_k$ & $\dot{\Pi}_k$ & 
Unstable\\ 
(s) & & (s) &  & & (s) & ($10^{-13}$ s/s) &  \\
\noalign{\smallskip}
\hline
\noalign{\smallskip}       
 423.246 & 1 &  420.975 & 1 &  8  &  2.272   & 0.725 & yes \\ 
 463.481 & 1 &  462.511 & 1 &  9  &  0.970   & 1.354 & yes \\
 464.238 & ? &  462.511 & 1 &  9  &  1.727   & 1.354 & yes \\
 494.026 & ? &  499.054 & 1 & 10  & $-5.028$ & 1.146 & yes \\
 538.300 & 1 &  538.126 & 1 & 11  &  0.174   & 1.394 & yes \\ 
 574.220 & 1 &  572.947 & 1 & 12  &  1.273   & 1.041 & yes \\
 618.272 & 1 &  617.647 & 1 & 13  &  0.625   & 1.319 & yes \\
 658.690 & 1 &  657.517 & 1 & 14  &  1.173   & 1.735 & yes \\ 
 699.820 & 1 &  696.097 & 1 & 15  &  3.723   & 1.650 & yes \\
 730.280 & 1 &  732.797 & 1 & 16  & $-2.517$ & 1.506 & yes \\
 775.745 & 1 &  770.251 & 1 & 17  &  5.494   & 1.621 & yes \\
 811.076 & 1 &  812.662 & 1 & 18  & $-1.586$ & 2.014 & yes \\
 854.593 & 1 &  853.775 & 1 & 19  &  0.818   & 2.026 & yes \\
 901.490 & ? &  894.617 & 2 & 36  &  6.873   & 2.176 & yes \\ 
 923.556 & ? &  926.073 & 1 & 21  & $-2.517$ & 1.900 & yes \\
 963.226 & ? &  965.149 & 2 & 39  & $-1.923$ & 2.704 & yes \\   
1014.350 & 1 & 1010.499 & 1 & 23  &  3.851   & 2.470 & yes \\    
1062.320 & ? & 1063.670 & 2 & 43  & $-1.350$ & 3.004 & yes \\
1087.538 & 1 & 1082.719 & 1 & 25  & 4.819    & 2.420 & yes \\
\noalign{\smallskip}
\hline
\end{tabular}
\label{table:GD358-asteroseismic-model}
\end{table*}

\begin{figure} 
\includegraphics[clip,width= 1.0\columnwidth]{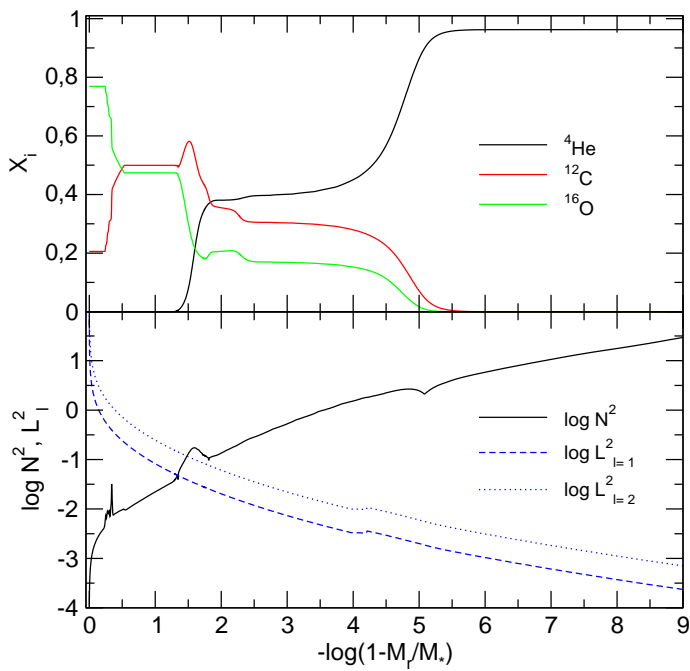}
\caption{Chemical  profiles  (upper panel)  and  the  squared 
Brunt-V\"aïs\"al\"a  and  Lamb  frequencies  for $\ell= 1$ and $\ell= 2$ (lower panel)  corresponding  to  our  asteroseismological  DB  WD  model  with  a  stellar  mass $M_{\star}= 0.584 M_{\odot}$ and an effective temperature $T_{\rm eff}= 24\,967 $ K.}
\label{fig:ast-model-gd358} 
\end{figure} 

We have also studied the pulsational stability/instability nature of the modes
associated with the periods fitted to the observed ones. 
We assume the frozen-in convection approximation \citep{1989nos..book.....U}. 
In particular,  we examine the sign 
and magnitude of the linear nonadiabatic growth rates, $\eta_k= -\Im(\sigma_k)/\Re(\sigma_k)$, 
where $\Re(\sigma_k)$ and $\Im(\sigma_k)$ are  the  real  and  the imaginary parts, respectively, of the complex eigenfrequency $\sigma_k$.  
A positive value of $\eta_k$ means that 
the mode  is linearly unstable (eighth column of Table 
\ref{table:GD358-asteroseismic-model}). In the left (right) hand panel of Fig. \ref{fig:per-teff-eft-0584-GD358}, we  show the  instability  domain 
of $\ell= 1$ ($\ell= 2$) periods as a function of the effective temperature for 
the DB WD model sequence with $M_{\star}= 0.584 M_{\odot}$. 
The palette of colors (right-hand scale) indicates the 
logarithm of the $e$-folding time, $\tau_{\rm e}$ (in years) of each excited mode, 
being $\tau_{\rm e}=1/|\Im(\sigma_k)|$. Many  pulsation  modes  
are excited, and the interval of periods corresponding to unstable modes 
of our asteroseismological model is nearly coincident with the range of  
the periods exhibited by GD~358 for most of the range of possible 
effective temperatures.
In particular, the pulsation periods of GD~358 fall into the highest 
excitation regimen (that is, shortest $e$-folding times), which reflects an excellent 
agreement between our nonadiabatic calculations and the observational data.

In Table~\ref{table:modelos-sismo-gd358}, we list the main
characteristics of GD~358 according to the previous studies and the present
work. Most of the data of this table are extracted from Table 7 of \cite{2019ApJ...871...13B}.
In the case of the results of the present work, the errors in $T_{\rm eff}$ and 
$\log(L_{\star}/L_{\odot})$ are estimated from the width of the maximum in 
the function $(\chi ^2)^{-1}$ vs $T_{\rm eff}$ and $\log(L_{\star}/L_{\odot})$,
respectively, and 
the error in the stellar mass comes from the grid resolution in $M_{\star}$. 
Errors in the remainder quantities, $\log g$ and $\log(R_{\star}/R_{\odot})$, 
are derived from these values. The seismological stellar mass  
($0.584^{+0.025}_{-0.19}M_{\odot}$) is somewhat larger than the values derived from spectroscopy, although still compatible with them ($0.508\pm 0.050 M_{\odot}$ and $0.560\pm 0.028M_{\odot}$) within their uncertainties. In addition, 
the stellar mass of the asteroseismological model is in excellent agreement
with the stellar mass value derived from the mean period spacing 
($0.588\pm0.024 M_{\odot}$). We can conclude that the three 
approaches to determine stellar mass of GD~358 give very similar results, which 
implies that it is a well-constrained quantity.  

\begin{table*}
\centering
\caption{The main characteristics of the DBV star GD~358. The second column  
corresponds to spectroscopic and astrometric results, whereas the third, fourth, fifth, sixth, seventh, and eighth
columns present the results from the seismological studies of BW94 \citep{1994ApJ...430..850B}, DK95 \citep{1995ApJ...445L.141D}, MEA00 \citep{2000ApJ...545..974M}, MEA01 \citep{2001ApJ...557.1021M}, FB02 
\citep{2002ApJ...581L..33F}, M03 \citep{2003ApJ...587L..43M}, and BKEA19 
\citep{2019ApJ...871...13B}, 
respectively, and the last column presents the seismological results from the present 
work.} 
\begin{tabular}{l|ccccccccc}
\hline
\hline
Quantity & Spectroscopy & BW94 & DK95 & MEA00 & MEA01 & FB02 & M03 & BKEA19 & Seismology  \\
         & Astrometry   &      &      &       &       &      &     &        & (this work) \\ 
\hline
$T_{\rm eff}$ [K]              & $24\,000\pm500^{(a)}$   & $24\,000$ & $24\,121$ & $22\,600$ & $22\,600$ & $24\,800$ & $22\,900$ & $23\,650$ & $24\,967\pm 200$ \\
                                & $24\,937\pm1018^{(b)}$ &           &           &           &           &           &           &           & \\ 
$\log g$ [cm/s$^2$]             &  $7.78\pm0.05^{(a)}$   & $8.0$     & $\cdots$      & $\cdots$      & $\cdots$      & $8.02$    & $\cdots$      &  $\cdots$     & $7.964^{+0.048}_{0.043}$\\
                                &  $7.92\pm0.05^{(b)}$   &           &           &           &           &           &           &           & \\    
$M_{\star}$ [$M_{\odot}$]       &  $0.508\pm0.050^{(c)}$  & $0.61$    & $0.58$    & $0.605$   & $0.65$    & $0.625$   & $0.66$    & $0.571$   & $0.584^{+0.025}_{-0.019}$\\   
                                &  $0.560\pm0.028^{(d)}$  &           &           &           &           &           &           &           & \\
$\log (L_{\star}/L_{\odot})$    &  $\cdots$                   & $-1.30$   & $\cdots$      & $\cdots$      & $\cdots$      & $-1.25$   & $\cdots$      & $-1.287$  & $-1.215\pm 0.015$\\   
$\log(R_{\star}/R_{\odot})$     &  $\cdots$              & $-1.90$   & $\cdots$      & $\cdots$      & $\cdots$      & $-1.89$   & $\cdots$      &  $\cdots$     & $-1.880\pm 0.014$\\    
$X_{\rm O}$ (center)            &  $\cdots$                   & 0.50      & 0.50      & 0.80      & 0.84      &  0.00     & 0.67      & 0.50      & 0.77    \\     
$\log(1-M_{\rm env}/M_{\star})$ &  $\cdots$                   & $\cdots$      & $-2.6$    & $-2.74$   & $-2.74$   & $-2.97$   & $\cdots$      & $-2$      & $-1.6$  \\    
$\log(1-M_{\rm He}/M_{\star})$  &  $\cdots$                   & $-5.70$   & $-6.0$    & $-5.97$   & $\cdots$      & $-5.80$   & $-2.0$      & $-5.5$    & $-5.98$ \\  
$d$  [pc]                       & $36.6\pm4.5^{(b)}$     & 42        & $\cdots$      & $\cdots$      & $\cdots$      &  43       & $\cdots$      & 44.5      & $42.85\pm 0.73$ \\   
                                & $43.02\pm0.04^{(e)}$ &           &           &           &           &           &           &           & \\   
\hline
\hline
\end{tabular}
\label{table:modelos-sismo-gd358}

{\footnotesize  References: 
(a)  \cite{2012ASPC..462..171N};  
(b) \cite{2017ApJ...848...11B}; 
(c) From the evolutionary tracks and the $T_{\rm eff}$ and $\log g$ values of \cite{2012ASPC..462..171N};
(d) From the evolutionary tracks and the $T_{\rm eff}$ and $\log g$ values of \cite{2017ApJ...848...11B};
(e) {\it Gaia}.}
\end{table*}

 In comparison with previous 
seismological studies, the effective temperature of our asteroseismological 
model is the largest one, although very similar to that derived by \cite{2002ApJ...581L..33F}, and in excellent agreement with the spectroscopic 
$T_{\rm eff}$ inferred by \cite{2017ApJ...848...11B}. Regarding the stellar mass 
of our asteroseismological model, its value is in excellent agreement with the 
values derived in all previous studies. Direct comparison of other quantities 
such as the central abundance of O ($X_{\rm O}$), the thickness of the envelope 
rich in O, C, and He, $\log(1-M_{\rm env}/M_{\star})$, or the thickness of the pure 
helium envelope, $\log(1-M_{\rm He}/M_{\star})$, becomes less clear because the 
chemical structure of our DB WD models is (by construction) substantially different 
from those used in previous studies. Even so, we can note  that 
the envelope of our asteroseismological model is somewhat thicker than in the
previous studies, and that the pure-He envelope has a thickness 
quite similar to the thickness derived in other works.

We succinctly describe  the  main  properties  of  our  asteroseismological 
DB model for GD~358. We display in Fig. \ref{fig:ast-model-gd358} the internal chemical profiles of  this model  (upper  panel),  where the abundance by mass of the main constituents ($^4$He, $^{12}$C, and $^{16}$O)  is  depicted  as a function  of  the  outer  mass  fraction  $[-\log(1-M_r/M_{\star})]$. The  chemical  structure  of  the  model  is characterized by a C/O core --resulting from the core He-burning phase of the 
prior evolution-- shaped by extra-mixing processes such as overshooting. The core is surrounded by a layer rich in He, C, and O, which results from the nucleosynthesis during the TP-AGB stage. Above this shell, there is a pure He mantle resulting from
gravitational settling that causes He to float to the surface and heavier species 
to sink. The lower panel of Fig. \ref{fig:ast-model-gd358} shows the squares of the two critical frequencies of nonradial stellar pulsations, that is,  the Brunt-V\"ais\"al\"a  frequency  and  the  Lamb  frequency  $L_{\ell}$ for
$\ell= 1$ and $\ell= 2$. The  shape  of  the  Brunt-V\"ais\"al\"a  frequency 
largely  defines the  properties of  the $g$-mode period spectrum of  the model. 
In particular, each chemical transition region in the model contributes locally to the 
value of $N$. The most notable characteristic is the highly peaked structure at the C/O chemical transition [$-\log(1-M_r/M_{\star}) \sim 0.34$]. On the other hand, there is  
the He/C/O interface  at $-\log(1-M_r/M_{\odot}) \sim 1.3-1.8$ that causes 
the presence of a notable bump in the Brunt-V\"ais\"al\"a frequency and affects the mode-trapping properties of the model.

\subsection{Asteroseismological distance}
\label{distance}

We can estimate an asteroseismological distance to GD~358 on the basis of 
the luminosity of the asteroseismological model 
[$\log(L_{\star}/L_{\odot})= -1.215\pm0.015$ and $T_{\rm eff}= 24\,967\pm200$ K]
and a bolometric correction $BC= -2.71$ \citep{1994ApJ...430..850B}. The absolute
magnitude can be assessed as $M_{\rm V}= M_{\rm B}-BC$,  where $M_{\rm
  B}= M_{{\rm B},\odot} - 2.5\ \log(L_{\star}/L_{\odot})$.  We employ
the solar bolometric magnitude $M_{\rm B \odot}= 4.74$
\citep{2000asqu.book.....C}. The seismological distance $d$  is
derived from the relation: $\log d= [m_{\rm V} - M_{\rm V} +5]/5$.
We use the apparent visual magnitude  $m_{\rm V}= 13.65 \pm 0.01$
\citep{1982ApJ...262L..11W},  and obtain the seismological distance
and parallax  $d= 42.85\pm 0.73$ pc and  $\pi= 23.33\pm 0.41$ mas. 
The uncertainty in the seismological distance comes
from the uncertainty in the luminosity of the
asteroseismological model. These values are consistent with the results of 
\cite{1994ApJ...430..850B} and \cite{2019ApJ...871...13B}, although a bit larger 
than the distance derived  by \cite{2017ApJ...848...11B}.
A very important check for the validation of the
asteroseismological model is the comparison of the
seismological  distance with the distance derived from astrometry. We
have available the estimates  from {\it Gaia} EDR3, $d_{\rm G}= 43.02 \pm 0.04$
pc and $\pi_{\rm G}= 23.244 \pm 0.024$~mas.  They are in excellent
agreement with the asteroseismological derivations. This adds 
robustness to the asteroseismological model we found for GD~358. Also, 
the match of seismological and trigonometric parallax confirm that we are seeing 
dipole modes.

\section{Summary and conclusions}
\label{conclusions}

 In this paper, we have presented new space observations of the already 
known DBV star 
GD~358.  This is the first time that this star is intensively examined 
by a space mission such {\it TESS}, which provides high-quality time-series photometry for asteroseismic purposes. The results of our analysis broadly confirm the previous observations from extensive
ground-based observational campaigns. We also carried out a detailed asteroseismological analysis employing 
fully evolutionary models of DB WDs. We find that the evolutionary and pulsational 
properties of GD~358 according to our analysis based on space data combined 
with ground-based data are in line with 
the results of previous asteroseismological analyzes of this star based on data from the ground alone. 
The present study is the third part of a series of 
papers devoted to the study of pulsating H-deficient
WDs observed with {\it TESS}. We extracted 26 periodicities including 8 combination frequencies from the {\it TESS} light curve of GD~358 using a standard pre-whitening procedure to derive the potential 
pulsation frequencies. The oscillation frequencies, associated to $g$-mode
pulsations, have periods from $\sim 422$ s to $\sim 1087$ s. We combined these space 
data with the abundant ground-based observations available and found a constant period 
spacing of $39.25\pm0.17$ s, which allowed us to infer its stellar mass 
($M_{\star}= 0.588\pm0.024 M_{\sun}$) and constrain the  harmonic degree $\ell$ of some 
of the modes. We performed a period-to-period fit analysis on  GD~358, which provides 
us with an asteroseismological model with a stellar mass ($M_{\star}= 0.584^{+0.025}_{-0.019} M_{\sun}$) 
in agreement with the stellar-mass value  inferred from the period spacing,  and 
also  compatible with the spectroscopic mass ($M_{\star}= 0.560\pm0.028 M_{\sun}$). 
In agreement with previous works, we found that the frequency splittings vary according to the 
radial order of the modes, suggesting differential rotation and preventing us to derive a 
reliable representative rotation period of the star. The seismological model derived from our analysis
allowed us to estimate the seismological distance ($d_{\rm seis}= 42.85 \pm 0.73$ pc) of 
GD~358, which is in excellent agreement with the precise astrometric  distance measured by 
{\it GAIA} EDR3 ($d_{\rm GAIA}= 43.02 \pm 0.04$ pc).

In accordance with the findings of our recent works focused on pulsating H-deficient WDs 
\citep[][]{2019A&A...632A..42B,2021A&A...655A..27U} 
we conclude that the high-quality data collected by the {\it TESS}
space mission,  combined with  ground-based photometric data, are able to 
provide a reliable input to the asteroseismology of WD stars. The {\it TESS} mission, 
along with  future space  missions and upcoming surveys, will allow an unprecedented 
boost to the stellar seismology of these ancient stars. 
  
\begin{acknowledgements}
Part of this work was supported by AGENCIA through the Programa de
  Modernizaci\'on Tecnol\'ogica BID 1728/OC-AR, and by the PIP
  112-200801-00940 grant from CONICET. 
  M.U. acknowledges financial support from CONICYT Doctorado Nacional in the form of grant number No: 21190886 and ESO studentship program.
  K.J.B. is supported by the National Science Foundation under Award AST-1903828.
  S.O.K.  acknowledges financial support from Coordena\c{c}\~ao de Aperfei\c{c}oamento de Pessoal de N\'{\i}vel Superior - Brasil (CAPES) - Finance Code 001, Conselho Nacional de Desenvolvimento Cient\'{\i}fico e Tecnol\'ogico - Brasil (CNPq), and Funda\c{c}\~ao de Amparo \`a Pesquisa do Rio Grande do Sul (FAPERGS) - Brasil.
  A.S.B. acknowledges financial support from the National Science Centre under projects No.\,UMO-2017/26/E/ST9/00703 and UMO-2017/25/B/ST9/02218.
  This research has made use of NASA's Astrophysics Data System. This paper includes data collected by the TESS mission. Funding for the TESS mission is provided by the NASA's Science Mission Directorate.

\end{acknowledgements}

\bibliographystyle{aa}
\bibliography{paper_bibliografia.bib}

\end{document}